%% file: main.tex
\documentclass[sigconf,screen]{acmart}
\AtBeginDocument{%
  \providecommand\BibTeX{{%
    \normalfont B\kern-0.5em{\scshape i\kern-0.25em b}\kern-0.8em\TeX}}}

\setcopyright{acmlicensed}
\copyrightyear{2024}
\acmYear{2024}
\acmDOI{10.1145/3658644.3690274}

\acmConference[CCS '24]{ACM Conference on Computer and Communications Security}{October 14--18,
  2024}{Salt Lake City, USA}
\acmISBN{978-1-4503-XXXX-X/18/06}

\input{misc/packages}
\input{misc/definitions}

\input{misc/commands}

\begin{document}

\title{No Peer, no Cry: Network Application Fuzzing via Fault Injection}

\author{Nils Bars}
\affiliation{
  \institution{CISPA Helmholtz Center for Information Security}
  \country{Germany}
}
\email{nils.bars@cispa.de}

\author{Moritz Schloegel}
\affiliation{
  \institution{CISPA Helmholtz Center for Information Security}
  \country{Germany}
}
\email{moritz.schloegel@cispa.de}

\author{Nico Schiller}
\affiliation{
  \institution{CISPA Helmholtz Center for Information Security}
  \country{Germany}
}
\email{nico.schiller@cispa.de}

\author{Lukas Bernhard}
\affiliation{
  \institution{CISPA Helmholtz Center for Information Security}
  \country{Germany}
}
\email{lukas.bernhard@cispa.de}

\author{Thorsten Holz}
\affiliation{
  \institution{CISPA Helmholtz Center for Information Security}
  \country{Germany}
}
\email{holz@cispa.de}

\input{content/00_abstract}


\setcopyright{none} %
\settopmatter{printacmref=false, printfolios=true,printacmref=true} %
\renewcommand\footnotetextcopyrightpermission[1]{} %
\pagestyle{plain}

\maketitle

\input{content/01_introduction}
\input{content/02_background}

\input{content/03_design}
\input{content/04_implementation}
\input{content/05_evaluation}
\input{content/06_discussion}

\input{content/07_related_work}

\input{content/08_conclusion}

\begin{acks}
We thank our anonymous reviewers for their valuable feedback. For their feedback on an earlier draft, we thank Tobias Scharnowski and Addison Crump. We would also like to express our gratitude to the team maintaining DCMTK and the F5 security incident response and development team at NGINX, particularly Jordan Zebor. Their quick and outgoing response during our responsible disclosure process and their thoughtful feedback on how to test their products have significantly helped us.

This work was funded by the European Research Council (ERC) under the consolidator grant RS$^3$ (101045669) and the German Federal Ministry of Education and Research (BMBF) under the grant CPSec (16KIS1899).
\end{acks}

\input{main.bbl}
\appendix

\input{content/A_appendix}

\end{document}

%% file: misc/packages.tex
\usepackage{xspace}

\usepackage{pifont}

\usepackage{multirow}

\usepackage{adjustbox}

\usepackage{siunitx}

\usepackage{balance}

\usepackage{paralist}

\usepackage{lipsum}  

\usepackage{transparent}

%% file: misc/definitions.tex
\definecolor{sangria}{rgb}{0.57, 0.0, 0.04}
\definecolor{pinegreen}{rgb}{0.0, 0.47, 0.44}
\definecolor{rossocorsa}{rgb}{0.83, 0.0, 0.0}
\definecolor{ao}{rgb}{0.0, 0.0, 1.0}
\definecolor{deepjunglegreen}{rgb}{0.0, 0.29, 0.29}
\definecolor{dartmouthgreen}{rgb}{0.05, 0.5, 0.06}

%% file: misc/commands.tex
\newcommand{\todo}[1]{} 

\newcommand{\fuzztructionnet}{\textsc{Fuzztruction-Net}\xspace}
\newcommand{\toolname}{\textsc{FT-Net}\xspace}

\newcommand{\numbugs}{23\xspace}
\newcommand{\numtargets}{16\xspace} %
\newcommand{\numuniquetargets}{eleven\xspace} %

\newcommand{\bugreportcve}[1]{\href{https://www.cve.org/CVERecord?id=#1}{\color{black}{#1}}}

\newcommand{\blackhref}[2]{\href{#1}{\color{black}{#2}}}

\newcommand{\generator}{weird peer\xspace}
\newcommand{\generatorsc}{Weird Peer\xspace}

\newcommand{\eg}{e.g.,\xspace}
\newcommand{\ie}{i.e.,\xspace}
\newcommand{\etal}{et~al.\@\xspace}
\newcommand{\cf}{cf.\@\xspace}

\newcommand{\fuzztruction}{\textsc{Fuzztruction}\xspace}
\newcommand{\aflpp}{\textsc{AFL++}\xspace}
\newcommand{\afl}{\textsc{AFL}\xspace}
\newcommand{\libfuzzer}{\textsc{LibFuzzer}\xspace}
\newcommand{\honggfuzz}{\textsc{Honggfuzz}\xspace}

\newcommand{\aflnet}{\textsc{AflNet}\xspace}
\newcommand{\sgfuzz}{\textsc{SGFuzz}\xspace}
\newcommand{\stateafl}{\textsc{StateAFL}\xspace}
\newcommand{\snipuzz}{\textsc{Snipuzz}\xspace}
\newcommand{\snapfuzz}{\textsc{Snapfuzz}\xspace}
\newcommand{\boofuzz}{\textsc{BooFuzz}\xspace}
\newcommand{\peach}{\textsc{Peach}\xspace}
\newcommand{\bleem}{\textsc{Bleem}\xspace}
\newcommand{\nyxnet}{\textsc{Nyx-Net}\xspace}
\newcommand{\ossfuzz}{\textsc{OSS-Fuzz}\xspace}

\newcommand{\client}{$^{\textrm{C}}$\xspace}
\newcommand{\server}{$^{\textrm{S}}$\xspace}

\newcommand{\dropbearServer}{\texttt{dropbear\server}\xspace}
\newcommand{\dropbearClient}{\texttt{dropbear\client}\xspace}

\newcommand{\opensslServer}{\texttt{openssl\server}\xspace}
\newcommand{\opensslClient}{\texttt{openssl\client}\xspace}

\newcommand{\libresslServer}{\texttt{libressl\server}\xspace}
\newcommand{\libresslClient}{\texttt{libressl\client}\xspace}

\newcommand{\gnutls}{\texttt{gnutls}\xspace}
\newcommand{\gnutlsServer}{\texttt{gnutls\server}\xspace}
\newcommand{\gnutlsClient}{\texttt{gnutls\client}\xspace}

\newcommand{\pjsipServer}{\texttt{pjsip\server}\xspace}
\newcommand{\pjsipClient}{\texttt{pjsip\client}\xspace}

\newcommand{\dcmtk}{\texttt{dcmtk}\xspace}
\newcommand{\dcmtkServer}{\texttt{dcmtk\server}\xspace}
\newcommand{\dcmtkClient}{\texttt{dcmtk\client}\xspace}

\newcommand{\sambaServer}{\texttt{samba\server}\xspace}
\newcommand{\sambaClient}{\texttt{samba\client}\xspace}

\newcommand{\mosquittoServer}{\texttt{mosquitto\server}\xspace}
\newcommand{\mosquittoClient}{\texttt{mosquitto\client}\xspace}

\newcommand{\livefivefivefive}{\texttt{live555}\xspace}
\newcommand{\livefivefivefiveServer}{\texttt{live555\server}\xspace}
\newcommand{\livefivefivefiveClient}{\texttt{live555\client}\xspace}

\newcommand{\curlClient}{\texttt{curl\client}\xspace}

\newcommand{\ngtcptwoClient}{\texttt{ngtcp2\client}\xspace}
\newcommand{\ngtcptwoServer}{\texttt{ngtcp2\server}\xspace}

\newcommand{\nginxServer}{\texttt{nginx\server}\xspace}

\newcommand{\opensshServer}{\texttt{openssh\server}\xspace}
\newcommand{\opensshClient}{\texttt{openssh\client}\xspace}

\newcommand{\apacheServer}{\texttt{apache\server}\xspace}

\newcommand{\llvmcov}{\texttt{llvm-cov}\xspace}

\renewcommand{\paragraph}[1]{\medskip\noindent\textbf{#1}\hspace{1ex minus .1ex}}

\newcommand{\circleone}{\ding{202}\xspace}
\newcommand{\circletwo}{\ding{203}\xspace}
\newcommand{\circlethree}{\ding{204}\xspace}
\newcommand{\circlefour}{\ding{205}\xspace}
\newcommand{\circlefive}{\ding{206}\xspace}

\newcommand{\cmark}{\textcolor{pinegreen}{\ding{51}}\xspace}
\newcommand{\xmark}{\textcolor{rossocorsa}{\ding{56}}\xspace}

%% file: content/00_abstract.tex
\begin{abstract}
Network-facing applications are commonly exposed to all kinds of attacks, especially when connected to the internet. As a result, web servers like Nginx or client applications such as curl make every effort to secure and harden their code to rule out memory safety violations. One would expect this to include regular fuzz testing, as fuzzing has proven to be one of the most successful approaches to uncovering bugs in software. Yet, surprisingly little research has focused on fuzzing network applications. When studying the underlying reasons, we find that the interactive nature of communication, its statefulness, and the protection of exchanged messages (\eg via encryption or cryptographic signatures) render typical fuzzers ineffective. Attempts to replay recorded messages or modify them on the fly only work for specific targets and often lead to early termination of communication.

In this paper, we discuss these challenges in detail, highlighting how the focus of existing work on protocol state space promises little relief. We propose a fundamentally different approach that relies on \emph{fault injection} rather than modifying messages. Effectively, we force one of the communication peers into a weird state where its output no longer matches the expectations of the target peer, potentially uncovering bugs. Importantly, this \emph{weird peer} can still properly encrypt/sign the protocol message, overcoming a fundamental challenge of current fuzzers. In effect, we leave the communication system intact but introduce small corruptions. Since we can turn either the server or the client into the weird peer, our approach is the first that can effectively test client-side network applications. In an extensive evaluation of \numtargets targets, we show that our prototype \fuzztructionnet significantly outperforms other fuzzers in terms of coverage and bugs found. Overall, \fuzztructionnet uncovered \numbugs new bugs in well-tested software, such as the web servers Nginx and Apache HTTPd and the OpenSSH client.

\medskip
\noindent{}\emph{This is the author's version of the work. It is posted here for your personal use. Not for redistribution. The definitive Version of Record was published in ACM Conference on Computer and Communications Security (CCS), \url{https://doi.org/10.1145/3658644.3690274}.}

\end{abstract}

%% file: content/01_introduction.tex
\section{Introduction}\label{sec:introduction}

The internet has defined many aspects of modern life, including the unhindered exchange of information, real-time communication across the globe, and the ability to conduct business entirely online. 
The very same connectivity that provides numerous benefits also introduces risks as internet traffic passes through multiple, potentially untrusted nodes. In effect, remote attackers can potentially exploit any bug or software fault to compromise systems, which emphasizes the importance of network-facing applications as a critical security boundary.

Even a single memory corruption error in the web server or network client can have disastrous consequences, which demands thorough testing of these applications.
To this end, we first need to understand the attack surface: %
Network applications receive and process messages that use a multitude of protocols to exchange data, such as HTTP, SSH, SIP, or SMB. 
These protocols define the rules and structures for communication and ensure that different devices can interpret and exchange data coherently.
Thus, automated testing must account for this structure, testing inputs that are valid to some degree, such that the application's parser does not immediately discard them but exercises deeper functionality.
Cryptographic techniques are widely used to secure communication, making these testing efforts more difficult. Protocols like TLS (Transport Layer Security) provide secure communication over networks that may not be inherently secure.
The transition to HTTP/3 underscores the importance of built-in encryption by incorporating the QUIC transport layer, which mandates TLS~1.3.
In addition to encryption, compression is used in many network protocols to reduce overhead and maintain efficiency, and checksums are commonly employed to ensure a reliable data transfer.
All these concepts complicate automated security testing, as these security measures make it difficult to generate valid messages.

Interestingly, the field of fuzzing, otherwise renowned for effective bug finding~\cite{ossfuzz}, shows a noticeable gap in testing network-facing applications.
Despite the advent of AFL~\cite{afl} sparking a renaissance of fuzzing methods with hundreds of new fuzzers proposed~\cite{aschermann2019redqueen,fioraldi2022libafl,aflplusplus,lyu2019mopt,yun2018qsym,boehme2020entropic,zhu2022survey}, most of them are limited to a specific subset of software.
Most benchmarks~\cite{metzman2021fuzzbench,hazimeh2020magma,google2016fts}, academic papers in general~\cite{schloegel2024sok}, and public industry initiatives such as OSS-Fuzz~\cite{ossfuzz} focus predominantly on comparably simple C/C++ Linux programs that consume byte-oriented input. %
In contrast, network applications are significantly more complex but have attracted relatively little attention in fuzzing research~\cite{daniele2024survey}. 

Broadly speaking, existing work can be split into two types of approaches. First, \emph{replay-based} fuzzers~\cite{pham2020aflnet,ba2022sgfuzz,natella2022stateafl} record valid communication and use saved messages as seeds. By replaying (and mutating) these messages, they do not require extensive protocol knowledge. However, this approach is limited, mainly because most protocols use ephemeral values, such as session identifiers that are dynamically created during runtime. 
These values prevent replaying old messages and lead to \emph{low input stability}, \ie executing the same input more than once will exercise entirely different code parts.
To avoid this problem, a second type of approach~\cite{luo2023bleem} uses a \emph{Man-in-the-Middle (MitM)} technique to modify messages in transit rather than replaying old ones. This way, session identifiers are implicitly adapted. However, this causes a state divergence between one peer sending the unmodified message and the other receiving a modified one. This leads to \emph{desynchronization}, where peers disagree on the current protocol state. Additionally, ubiquitous integrity checks or encryption prevent modification of messages in transit, as traditional fuzzer mutations immediately invalidate the integrity. The receiving peer then discards such invalidated messages before any interesting functionality is exercised. 
These challenges limit the effectiveness of existing work for many network applications, confining them to a small subset of applications that feature no integrity protection or ephemeral values. %
Instead of addressing these challenges, we find that state-of-the-art fuzzers focus on \emph{protocol state coverage}. This metric measures the extent to which the fuzzer has explored the various states defined by the protocol's state machine as an additional feedback mechanism~\cite{natella2022stateafl,zhao2022statefuzz,pham2020aflnet}. 
While this may be beneficial to navigate the intricacies of the protocol's logic, 
it implicitly assumes that the fuzzer can successfully exchange multiple messages without desynchronization and mutate messages without invalidating them---to the best of our knowledge, this is not the case for any state-of-the-art network application fuzzer.

In this paper, we present the design and implementation of \fuzztructionnet, a powerful and flexible approach to network application fuzzing.
Instead of replacing any peer or the communication channel, replaying observed messages, or modifying messages as MitM, our method uses one of the communication endpoints (\ie either client or server) to generate protocol messages. 
This approach enables testing protocols that rely on continuous exchange and stateful interactions. Ephemeral values or encrypted messages pose no problem, as our mutations reside within the communication system.
More precisely, \fuzztructionnet uses high-level mutation strategies based on injecting faults, a method previously explored in other contexts~\cite{bars2023fuzztruction, liu2021ifizz, jiang2020fifuzz,sharma2024fuzzerr}, into the endpoint that does not represent the system under test. 
These strategies include manipulating the application's control flow, such as flipping branching decisions or changing function call destinations to explore different execution paths and protocol states. 
This allows our approach to mutate the generation process of messages (rather than generated messages) while still preserving their semantics, effectively enabling us to explore deeper application logic.
A compelling aspect of \fuzztructionnet is its ability to handle common obstacles such as cryptographic operations, compression algorithms, and checksums, as faults can be injected before protocol messages are wrapped by these complex primitives. 
Moreover, our approach enables the fuzzing of both client and server components by switching the endpoint targeted for fault injection, a significant advantage over traditional methods focusing exclusively on server-side fuzzing.

We implemented a prototype of \fuzztructionnet and found that it significantly outperforms existing state-of-the-art fuzzers, finding on average 16\% more coverage and uncovering 3x more bugs than the second-best fuzzer, \sgfuzz. Also, \fuzztructionnet finds bugs in Nginx, the OpenSSH client, and apache2 that previous fuzzing attempts failed to uncover despite existing fuzzing harnesses and OSS-Fuzz integration.

\paragraph{Contributions.} We make the following main contributions:
\begin{itemize}
    \item We systematically analyze the shortcomings of state-of-the-art network application fuzzers, finding that they address neither desynchronization nor input stability.
    \item To address these challenges, we present \fuzztructionnet, our approach that uses fault injection to turn one communication partner into a weird peer that can be used for testing.
    \fuzztructionnet is the first network application fuzzer that can test both servers \emph{and} clients, compared to existing methods that are limited to servers only.
    \item We found \numbugs remotely triggerable bugs in popular network applications, such as the Nginx web server, the OpenSSH client, and the popular QUIC library ngtcp2 used by curl.

\end{itemize}

We release our prototype at \url{https://github.com/fuzztruction/fuzztruction-net} upon acceptance and intend to participate in the artifact evaluation process.

%% file: content/02_background.tex
\section{Background \& Challenges}\label{sec:background}
Before diving into our approach towards fuzzing network applications, we briefly discuss how network applications operate, enumerate the challenges they pose for fuzzing, and discuss how state-of-the-art approaches attempt to tackle the problem.

\subsection{Network Applications}
When referring to network applications, we consider a scenario where multiple applications, called peers, exchange data.
Typically, one of these peers takes the role of a \emph{server}, responsible for serving the requests of one or multiple \emph{clients}. 
The server continuously listens on a specified IP address and port, waiting for the clients to initiate a connection. 
To process multiple requests in parallel, the server-side software is either multithreaded or uses an event-driven design based on synchronous I/O multiplexing as provided by the \texttt{select} syscall and alike.
Sometimes, as for WebRTC-based video calls, no peer takes a distinct role; instead, clients communicate directly with each other. This work focuses on the more prevalent client-server architecture. %

Regardless of the setup, (interactive) communication takes place across a network. To this end, a transport protocol based on the IP protocol is usually used to exchange messages. The most common protocols are TCP and UDP, which significantly differ in reliability, logical connection type, and data flow properties. Notably, some applications employ their own transport protocol, which can be based on existing ones or entirely independent of commonly used protocols. One example of such protocols is QUIC, which is based on UDP and is intended to replace TCP with better performance properties and native support for encryption and authentication while providing the same reliability as TCP. 

Generally speaking, network applications significantly contrast traditional fuzzing targets, where we execute only a single application (\eg a binary executable). Such targets usually do not interact with other applications, and their execution is---nondeterminism aside---mainly determined by the given input.

\subsection{Fuzzing Challenges}%
\label{sec:challenges}
The differences between network applications and typical targets lead to three key challenges that must be addressed by any fuzzing approach to effectively and efficiently test network applications.

\paragraph{\bfseries C1. Session state}
To facilitate \emph{meaningful} communication, the client and the server maintain state throughout their communication. Crucially, some parts of the state are \emph{ephemeral}, such as session identifiers. This has a significant impact on the fuzzing process. An input that achieved high coverage in the first run may fail in the second run \emph{unless} the temporary identifier is updated to the new one. The server may send and expect to receive these identifiers at runtime, challenging the fuzzer's ability to predict them in advance. One famous example of such behavior is HTTP digest authentication~\cite{rfc7616}, where a server-chosen nonce protects communication against replay attacks. 
These dynamically chosen ephemeral values lead to the problem of \emph{low input stability}, as inputs cannot be replayed.
At the same time, failure to manage the session state across roundtrips leads to \emph{desynchronization}, where the peers disagree on the current protocol state. %

\paragraph{\bfseries C2. Integrity checks and encryption}
To protect messages, server and client use various integrity checks, including encodings, checksums, and encryption. Modifying these messages is virtually impossible for typical fuzzer mutations, as local changes invalidate the message integrity. Consequently, the recipient discards the message during parsing without executing more interesting code parts.
Naively, testing can record live communication and replay these messages.
However, modern cryptographic algorithms require the use of ephemeral values to protect against codebook attacks, mandating to solve challenge C1.

\paragraph{\bfseries C3. Network Application Setup}
Assuming a fuzzer can maintain state across iterations and has found a way to deal with any integrity checks or even encryption, it must still be tailored towards the system under test (SUT). In particular, this includes aspects such as when to send fuzzing input---as a server may need to perform initialization before accepting connections---or when to terminate the fuzzing session---as a server may wait for additional client messages rather than terminating itself, contrasting file processing applications, which usually exit after having consumed all input. 

\subsection{Modern Network Application Fuzzing}
Several approaches have been proposed to fuzz test network applications, and we will summarize the state-of-the-art work next.

\paragraph{\bfseries \aflnet~\cite{pham2020aflnet}.}
\aflnet is the first grey box fuzzer for network protocol implementations based on \afl. 
It focuses on integrating protocol state feedback into \afl's coverage-guided fuzzing approach.
Central to \aflnet's approach is the isolation of \emph{protocol data units} (PDUs), \ie messages as defined by the underlying protocol. Using these isolated messages as building blocks, \aflnet navigates the protocol state space by permuting the order of messages or by mutating their content. To infer the already covered protocol state space, \aflnet uses manually crafted parsers to assign an identifier to received messages. In essence, the order of received messages and their assigned identifier determines the path exercised in the inferred state machine.
Since \aflnet is based on \afl, it inherits its limitations regarding ephemeral values (C1) and any kind of integrity protection (C2). Also, a human expert is required to create a parser for the target protocol to infer the protocol state from messages.

\paragraph{\bfseries \sgfuzz~\cite{ba2022sgfuzz}.}
\sgfuzz is an in-process fuzzing tool based on \libfuzzer and uses a library provided by \honggfuzz~\cite{honggfuzz} to enable the fuzzing of network applications. The core design of \sgfuzz assumes that the current state of the protocol is reflected in the runtime values of \texttt{enum} variables.
To monitor these variables, the source code of the target application is preprocessed. This involves wrapping the assignment expressions of enum-typed variables (\ie \verb|enum_t v = <enum_variant>|), enabling observation of the assigned values. A \emph{state transition tree} is constructed from the values observed during fuzzing. %
If new transitions in the protocol state are discovered, \sgfuzz can use this information to determine if a new protocol state has been encountered.
Like \aflnet, \sgfuzz entirely relies on the fuzzing logic provided by its baseline, \ie \libfuzzer, which features bit-level mutations incapable of overcoming more complex fuzzing roadblocks (C1, C2). Furthermore, the \honggfuzz library used for network fuzzing does not support UDP, limiting \sgfuzz' applicability to specific protocol implementations.

\paragraph{\bfseries \stateafl~\cite{natella2022stateafl}.}
Based on \aflnet, \stateafl uses a different approach than its baseline to tackle the problem of protocol state space exploration. To avoid manual effort, \stateafl uses the content of memory allocations to infer the current protocol state. By intercepting APIs related to networking, such as those creating sockets or those reading or writing data to them, \stateafl infers possible points in time at which it checks if the content of tracked memory objects indicates that a new state has been covered.
This approach comes with the risk of introducing errors into the target under test, \eg if the memory tracking is unsound or incomplete, the instrumentation may access a memory object that is not allocated. Since \stateafl is based on \aflnet, it suffers from the same limitations regarding more complex targets that rely on session state (C1) or cryptography (C2).

\paragraph{\bfseries \bleem~\cite{luo2023bleem}.}
The recently proposed approach called \bleem does \emph{not} replace the client with a fuzzer---contrasting previous approaches---but instead operates as MitM.
It connects to both communication parties and forwards messages between them. This setup uses protocol parsers to deduce the protocol's current state, which helps construct a graph representing the system's state. Using this graph, \bleem generates new packet sequences or mutates the transmitted data.
While this MitM approach addresses the input stability problem of C1 and allows \bleem, for example, to perform a TLS handshake by simply forwarding messages without modifications, this may still cause desynchronization, as the sent and received messages differ, causing different states in each peer. Also, \bleem lacks the ability to mutate encrypted or integrity-protected data (C2) and requires hand-crafted protocol parsers to infer protocol state.

\medskip
All state-of-the-art fuzzers using a replay-based approach fail to address C1 and C2. Only \bleem, with its MitM-based approach, overcomes C1 partly but still does not account for integrity protection. All fuzzers tailor their approach towards network application fuzzing, addressing C3, and use different forms of protocol state as feedback for guiding the fuzzer. However, without solving C1 and C2, their approaches are severely limited and require patching of targets (\eg to remove integrity protection) for any meaningful exploration.

%% file: content/03_design.tex
\section{Design}\label{sec:design}
To effectively and efficiently test network applications, we need a design that maintains session state across exchanged messages (C1), accounts for integrity protection mechanisms (C2), and is tailored towards the needs of network applications (C3).

\subsection{Idea \& Rationale} A fuzzer intending to address these challenges needs to adapt its machinery to take previously exchanged messages into account. The complexity and diversity of network applications make this problem hard to solve in the general case: Without re-implementing all protocols, a fuzzer's heuristics will always fall short in some regard. We argue that replacing one peer with a fuzzer, as done by \aflnet, \sgfuzz, or \stateafl, is not only unnecessary but potentially limiting successful testing. The fuzzer taking the peer's place has no notion of state to preserve across exchanged messages, nor does it know the underlying protocol or state machine. Thus, discarding the original peer with all its domain knowledge potentially degrades the system's quality regarding communication breadth and depth.
Furthermore, we argue that placing the fuzzer as MitM, as done by \bleem, has similar shortcomings. In a sense, this approach replaces the communication channel with a fuzzer-controlled one. While leveraging the domain knowledge inherent in a peer, any communication protected by integrity checks or encryption is out of reach, limiting the applicability of this approach to specific network protocols. Even if messages are not protected, any modifications made by the fuzzer will only be known to the recipient and not the sender. This leads to \emph{peer desynchronization}, as one peer's internal state diverges from the other's.

In this work, we propose a method to introduce a fuzzer into this communication system while replacing \emph{neither} any peer \emph{nor} the communication channel. This way, we can preserve the dynamic nature of the communication between peers and maintain synchronicity of the protocol state on both sides. Naturally, these requirements leave us with little room for where to place our fuzzer. Especially the second constraint, \ie preventing desynchronization, mandates including the fuzzer in one peer itself, albeit without replacing it entirely.
Fundamentally, we rely on \emph{fault injection}~\cite{bars2023fuzztruction, liu2021ifizz, jiang2020fifuzz} to achieve our goal. Rather than mutating specific input bytes, we mutate the peer itself, causing its output to produce unexpected or---at times---incorrect messages. More specifically, we propose injecting faults into the peer's state-modifying operations, \eg we could inject a fault into the selection of an encryption algorithm, provoking an untypical choice the other peer may not expect. At the same time, we need to avoid injecting faults into the encryption algorithm itself, as this would invalidate the produced output, which is unlikely to pass the other peer's decryption stage. 
In essence, our faults should neither invalidate the communication nor cause an unsynchronized protocol state transition, leading to the peers' disagreeing on the current state and, thus, terminating communication. 
If we can achieve these design goals, our fuzzer can play off one peer against the other.
We stress that this enables us to fuzz one peer \emph{or} the other, \ie we could target the server, as all existing work does, or the client, which existing tools largely neglect. 

\begin{figure}[tp]
    \centering
    \graphicspath{{graphics}}
    \def\svgwidth{\columnwidth}
    \begin{scriptsize}
        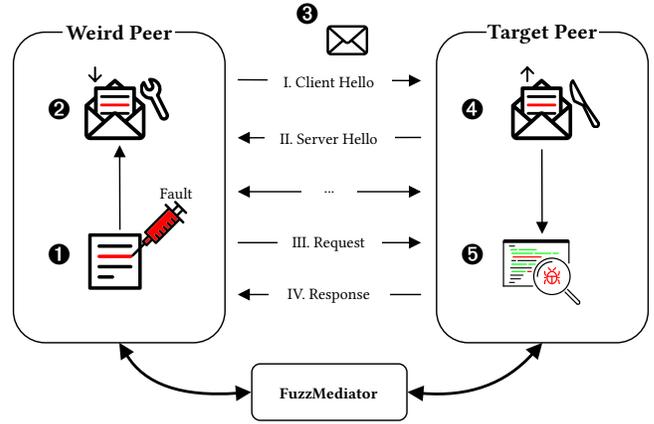
    \end{scriptsize}
    \vspace{-1.5em}
    \caption{High-level overview of \fuzztructionnet. Notably, the role of weird and target peer is independent of which peer is acting as server or client.}
    \label{fig:system_overview}
\end{figure}

\subsection{Overview}

Figure~\ref{fig:system_overview} shows an overview of our proposed design and sketches the exemplary interaction of two peers. For the sake of this example, we assume our goal is fuzzing the server, \ie the server is the \emph{target peer}, while we inject faults into the client, making it the \emph{weird peer} (in reference to \emph{weird machine}~\cite{Bratus2011weird}). We stress that the inverse setup is possible as well.
Beyond these two communicating peers, we use an external component called \emph{FuzzMediator} that manages the fuzzing loop by observing and controlling the other components of the system. The fuzzing loop starts with \circleone injecting some fault in the \emph{weird peer} that leads to unexpected behavior, \eg during the preparation of a protocol message. Ideally, the fault corrupts the message slightly, but the subsequent logic still processes it normally, \eg encrypting it as expected by the communication protocol (\circletwo). %
Then, \circlethree the client initiates communication with the server, \eg by sending a Client Hello message (I.). Upon receiving the message, the server decodes it (\circlefour) before further processing, potentially triggering a bug (\circlefive) due to the corruption. If no bug leads to an abnormal termination, the communication (and thus the current fuzzing iteration) continues as the server responds to the received message. The weird peer then processes this response (II.). 
At this point, we can inject another fault to further alter the expected message flow or content of the following message (III.). Once the message is sent, the server will again respond (IV.), continuing this communication loop until one of the peers terminates (normally or abnormally). This concludes one fuzzing iteration, after which we collect coverage information of the target peer. Collected code coverage is reported to the \emph{FuzzMediator}, which can be used as feedback to decide if one (or more) injected fault(s) lead to interesting behavior and schedule the next fuzzing iteration. 

\subsection{Fault Injection}%
\label{sec:design:fault_injection}

Now that we have outlined our design goals and presented an overview, we focus on the \emph{fault injection}, the centerpiece that turns one of the peers into the \emph{weird peer}. This step is the key ingredient towards introducing some error into the regular communication flow that then, in turn, uncovers a bug in the \emph{target peer}. In principle, our goal is to corrupt message content to some minor extent without modifying its later processing, such as calculating a checksum or encrypting it. This may uncover incorrect or missing validation in the other peer without the message being easily discarded due to violating the integrity protection mechanisms. In our design, this can be achieved easily \emph{if} we know what part of the program to mutate and in which way. However, we lack information on both \emph{where} and \emph{how} to mutate the program. To address either, we rely on the inherent randomness in the fuzzing process.

\paragraph{Fault Types.} Before focusing on the \emph{where}, let us discuss the faults we can inject by mutating the weird peer. In particular, we target the following instructions to inject different fault types:
\begin{itemize}
    \item Load/stores: modify \emph{values} loaded from/stored to memory.
    \item Switch case: change the \texttt{case} statement executed.
    \item Conditions: invert conditional moves or branches. %
    \item Calls: change the \texttt{call} destination or skip a call.
\end{itemize}

These faults operate at two levels: the \emph{value} level, where we can modify data values, and the \emph{control-flow} level, where we can invert a conditional branch or even change the called functions.
The former helps corrupt data values, while the latter may execute unexpected code, leading to \emph{weird} data processing. 
For fine-grained control over the particular fault, each location receives a fuzzer-generated input stream, which we can use to inject a different fault each time this location is executed. The length of this stream must account for the \emph{fault type}, \eg inverting a condition only requires one bit as opposed to modifying a 64-bit store, which requires eight bytes, and it needs to account for the number of times the location is executed. 
During each execution, we extract the necessary bytes from the input stream to XOR with the loaded/stored value, influence a condition's outcome, or affect which function is called.

\paragraph{Fault Locations.}
Now, as the instructions at which we can inject a fault occur thousands of times in network applications, the core question is \emph{where} to introduce our faults.
Fundamentally, we can address this by (1) trying to obtain more information using static analysis or similar approaches or (2) randomly introducing faults, pruning ineffective ones, and focusing on high throughput. Sticking to the fuzzing spirit, where high performance has been shown to trump informed but costly analyses~\cite{aschermann2019redqueen}, we focus on the second option. 

Consequently, we mark all the instructions mentioned above during instrumentation at compile-time, leaving us with many potential targets for fault injection. To improve our odds of injecting faults at ``good'' locations, our fuzzing campaign features a dedicated \emph{queue initialization}, which we describe subsequently when presenting the details of our fuzzing loop. 

Despite our efforts, some fault locations will inevitably be less effective due to the random nature of the injection process. For example, faults could invalidate ephemeral value handling and encryption or lead to protocol state desynchronization. In such cases, the communication between the weird and target peer cannot proceed as desired, resulting in an early termination of the process. This early exit automatically leads to less coverage being exercised, so the fuzzer quickly avoids scheduling these faults again. Consequently, few computational resources are spent on these less effective faults.
The same principle applies to \emph{redundant} fault locations. By leveraging coverage to filter out ineffective locations, we can utilize the weird peer's capabilities in handling ephemeral values and managing state throughout the communication without needing to identify or model these capabilities explicitly.

\subsection{Fuzzing Loop}

Our fuzzing loop follows roughly those of regular fuzzing methods. The major difference is that one queue entry consists of a list of tuples \texttt{[(loc, stream), ...]}. \texttt{loc} denotes the fault injection location at which we apply the byte \texttt{stream} to mutate the value, condition, or call destination. 

\paragraph{\bfseries Calibration.}
As typical AFL-based fuzzers do, we run a calibration phase for each new fuzzing queue entry. In our case, we measure how often each fault injection location is hit, allowing us to calculate how much fuzzing input is needed during the execution of this queue entry. For example, a branch consumes 1 bit of fuzzing input to decide if it should be flipped. Thus, we need one bit of input each time one specific branch is executed.

\paragraph{\bfseries Queue Initialization.}
Initially, our queue is empty. Due to the high number of fault locations, we initialize the queue as follows: We test a few mutations for each location and observe the target peer. If any of the mutations lead to new coverage, we create a new queue entry consisting of the tuple \texttt{[(loc, stream)]}. On the other hand, if the mutations repeatedly caused the weird peer to crash, we skip this location. This way, we automatically bias the fuzzer towards the initially successful fault locations, as the fuzzer's scheduler primarily picks entries from the queue.

\paragraph{\bfseries Fuzzing Iteration \& Mutations.}
From the fuzzer's perspective, one fuzzing loop consists of the following steps: \begin{inparaenum}\item Pick one queue entry, \item mutate this queue entry, \item execute the peers, and \item collect exercised coverage. \end{inparaenum}
To mutate queue entries, we can:
\begin{itemize}
    
    \item Mutate the \texttt{stream}: We mutate the applied \texttt{stream} using typical fuzzer mutations, such as bitflips or havoc. This, for example, changes the value loaded at the fault location or leads to the inversion of the branch condition.
    \item Splice queue entries: We can append a tuple from a second queue entry to the list of the current one. This effectively leads to combining two (or more) different faults.
    \item Extend queue entry: We can also attempt to append a newly generated tuple to the current queue entry, \ie we try to find another interesting fault by selecting an unused fault location and generating a \texttt{stream} using the same mutation as in the first step.
\end{itemize}

\subsection{Network Application Specifics}

Given the nature of network communication, our design must address its unique needs and requirements, as discussed next. %

\paragraph{\bfseries Temporal Input Channel.}
The server must be running to ensure that the client can perform any meaningful action. Therefore, ensuring the client and server have the correct startup order is crucial.
While for traditional fuzzing, it is sufficient to pass the fuzzing input via a file that is consumed as soon as the target is ready, the process is slightly more involved for networked targets.
One common technique to determine whether the server is ready to process input sent by a client is polling, \eg as used by \aflnet. This is facilitated by consecutively trying to establish a connection to the server's port until it is successful or some timeout is hit. This comes at the cost of wasting CPU cycles and cannot deal with ``connectionless'' protocols such as UDP, where no handshake occurs that would indicate whether the target port is active.

We rely on the server to tell us when it is ready to accept connections to avoid these shortcomings. We achieve this by hooking two functions: \texttt{listen} and \texttt{bind}. When one of these functions is called, the server can notify the \emph{FuzzMediator} that it is ready to accept client requests. This requires no active waiting and works for all protocols.

\paragraph{\bfseries Termination Point.}
Server applications typically run indefinitely while consecutively serving clients. Thus, when fuzzing a server, the state is possibly carried over between multiple executions, \ie client connections. This shares the same downsides as in-process fuzzing (\eg the observed behavior is not necessarily reproducible).
Therefore, server applications are typically patched such that they terminate after handling a single client connection. This approach works reliably for connection-oriented protocols since the fuzzing iteration is terminated when the server closes the client connection and the client closes the connection, which the server can observe due to its connection-oriented nature.
Unfortunately, such patching mechanisms do not work well for targets based on connectionless protocols since the server may wait for a client that already exited. Since the connection between client and server has no state, the server cannot detect such a situation except by using timeouts as an indicator. Similar to other approaches~\cite{pham2020aflnet,natella2022stateafl}, we tackle this problem by terminating the target via a signal after each fuzzing iteration if the target peer has not terminated earlier.

%% file: graphics/fuzztruction_design_v6.4.pdf_tex
\begingroup%
  \makeatletter%
  \providecommand\color[2][]{%
    \errmessage{(Inkscape) Color is used for the text in Inkscape, but the package 'color.sty' is not loaded}%
    \renewcommand\color[2][]{}%
  }%
  \providecommand\transparent[1]{%
    \errmessage{(Inkscape) Transparency is used (non-zero) for the text in Inkscape, but the package 'transparent.sty' is not loaded}%
    \renewcommand\transparent[1]{}%
  }%
  \providecommand\rotatebox[2]{#2}%
  \newcommand*\fsize{\dimexpr\f@size pt\relax}%
  \newcommand*\lineheight[1]{\fontsize{\fsize}{#1\fsize}\selectfont}%
  \ifx\svgwidth\undefined%
    \setlength{\unitlength}{339bp}%
    \ifx\svgscale\undefined%
      \relax%
    \else%
      \setlength{\unitlength}{\unitlength * \real{\svgscale}}%
    \fi%
  \else%
    \setlength{\unitlength}{\svgwidth}%
  \fi%
  \global\let\svgwidth\undefined%
  \global\let\svgscale\undefined%
  \makeatother%
  \begin{picture}(1,0.65486726)%
    \lineheight{1}%
    \setlength\tabcolsep{0pt}%
    \put(0,0){\includegraphics[width=\unitlength,page=1]{fuzztruction_design_v6.4.pdf}}%
    \put(0.16690266,0.59955752){\color[rgb]{0,0,0}\makebox(0,0)[t]{\lineheight{1.25}\smash{\begin{tabular}[t]{c}\small\textbf{Weird Peer}\end{tabular}}}}%
    \put(0.07300885,0.24557522){\color[rgb]{0,0,0}\makebox(0,0)[t]{\lineheight{1.25}\smash{\begin{tabular}[t]{c}\Large\circleone\end{tabular}}}}%
    \put(0.07300885,0.47345133){\color[rgb]{0,0,0}\makebox(0,0)[t]{\lineheight{1.25}\smash{\begin{tabular}[t]{c}\Large\circletwo\end{tabular}}}}%
    \put(0,0){\includegraphics[width=\unitlength,page=4]{fuzztruction_design_v6.4.pdf}}%
    \put(0.83064384,0.60083457){\color[rgb]{0,0,0}\makebox(0,0)[t]{\lineheight{1.25}\smash{\begin{tabular}[t]{c}\small\textbf{Target Peer}\end{tabular}}}}%
    \put(0.72123894,0.24557522){\color[rgb]{0,0,0}\makebox(0,0)[t]{\lineheight{1.25}\smash{\begin{tabular}[t]{c}\Large\circlefive\end{tabular}}}}%
    \put(0.72123894,0.47345133){\color[rgb]{0,0,0}\makebox(0,0)[t]{\lineheight{1.25}\smash{\begin{tabular}[t]{c}\Large\circlefour\end{tabular}}}}%
    \put(0,0){\includegraphics[width=\unitlength,page=7]{fuzztruction_design_v6.4.pdf}}%
    \put(0.49564924,0.0359823){\color[rgb]{0,0,0}\makebox(0,0)[t]{\lineheight{1.25}\smash{\begin{tabular}[t]{c}\textbf{FuzzMediator}\end{tabular}}}}%
    \put(0,0){\includegraphics[width=\unitlength,page=8]{fuzztruction_design_v6.4.pdf}}%
    \put(0.49567578,0.27280752){\color[rgb]{0,0,0}\makebox(0,0)[t]{\lineheight{1.25}\smash{\begin{tabular}[t]{c}III. Request\end{tabular}}}}%
    \put(0,0){\includegraphics[width=\unitlength,page=9]{fuzztruction_design_v6.4.pdf}}%
    \put(0.49592799,0.19150221){\color[rgb]{0,0,0}\makebox(0,0)[t]{\lineheight{1.25}\smash{\begin{tabular}[t]{c}IV. Response\end{tabular}}}}%
    \put(0,0){\includegraphics[width=\unitlength,page=10]{fuzztruction_design_v6.4.pdf}}%
    \put(0.49596782,0.52405849){\color[rgb]{0,0,0}\makebox(0,0)[t]{\lineheight{1.25}\smash{\begin{tabular}[t]{c}I. Client Hello\end{tabular}}}}%
    \put(0,0){\includegraphics[width=\unitlength,page=11]{fuzztruction_design_v6.4.pdf}}%
    \put(0.49596782,0.43447345){\color[rgb]{0,0,0}\makebox(0,0)[t]{\lineheight{1.25}\smash{\begin{tabular}[t]{c}II. Server Hello\end{tabular}}}}%
    \put(0,0){\includegraphics[width=\unitlength,page=12]{fuzztruction_design_v6.4.pdf}}%
    \put(0.46238938,0.62389381){\color[rgb]{0,0,0}\makebox(0,0)[t]{\lineheight{1.25}\smash{\begin{tabular}[t]{c}\Large\circlethree\end{tabular}}}}%
    \put(0,0){\includegraphics[width=\unitlength,page=13]{fuzztruction_design_v6.4.pdf}}%
    \put(0.49668463,0.35812169){\color[rgb]{0,0,0}\makebox(0,0)[t]{\lineheight{1.25}\smash{\begin{tabular}[t]{c}...\end{tabular}}}}%
    \put(0,0){\includegraphics[width=\unitlength,page=14]{fuzztruction_design_v6.4.pdf}}%
    \put(0.25663717,0.34955752){\color[rgb]{0,0,0}\makebox(0,0)[t]{\lineheight{1.25}\smash{\begin{tabular}[t]{c}Fault\end{tabular}}}}%
    \put(0,0){\includegraphics[width=\unitlength,page=15]{fuzztruction_design_v6.4.pdf}}%
  \end{picture}%
\endgroup%

%% file: content/04_implementation.tex
\section{Implementation}\label{sec:implementation}

We implement a prototype of \fuzztructionnet based on the fault injection framework offered by \fuzztruction~\cite{bars2023fuzztruction}. In particular, we inherit its use of LLVM's stack maps and patch points to dynamically mutate \texttt{store} and \texttt{load} operations in a program.
We have extended the framework to suit our needs and made several key modifications, which we discuss in the following. We implemented about $7,900$ lines of Rust and $800$ lines of C/C++ code to build \fuzztructionnet on top of \fuzztruction. %

\paragraph{Fault Injection Mechanism.}
Despite being conceptually similar in terms of injecting faults, we had to revamp Fuzztruction's way of mutating LLVM's live values, which is used to apply faults at desired locations.
The existing framework uses a so-called stack map to track values it wants to mutate. At runtime, it would then identify the register holding this value and---using a JIT compiler---create a function to mutate this value. 
While effective for modifying ``direct'' values that are loaded or stored, this approach does not work for \emph{conditions}, which are managed via the \texttt{EFLAGS} register: The compiler assumes that the function created by the JIT compiler clobbers the register and will evaluate the condition only \emph{after} our JIT-ed function call, making it impossible to modify the condition.
Therefore, we implemented a completely new fault injection strategy and replaced the existing one:
\begin{inparaenum}
\item We spill values that we want to mutate onto the stack, %
\item we track the spill locations in a stack map, %
\item and if mutating a condition, we spill the condition onto the stack before using a JIT-compiled function to mutate the value directly on the stack.
\item After the function mutated the value, we read it back from the stack and use it, \eg in the \texttt{cmp} instruction.
\end{inparaenum}

\paragraph{Function Call Injection.}
We also incorporate a new method of injecting faults that allows us to call another function from the entry of an executed function. 
To achieve this, for each function, we generate a function pointer table during instrumentation that tracks functions with the same signature (\ie same argument number and types). Additionally, we insert a patch point at function entry points that---when selected as a fault injection site---consumes one byte of fuzzing input that serves as an index into the function pointer table and redirects the control flow to this function instead. 

\paragraph{Network Applications Support.}
The \fuzztruction framework is by design incompatible with network targets because it expects its targets to be executed once and sequentially. We reworked the framework to support the concurrent and repetitive execution of targets, mimicking the interactive nature of network communication. Furthermore, we added support for network namespaces, which allow each application to have a distinct set of network interfaces. This allows server applications to run on the same port in parallel, a necessary precondition for multiple parallel fuzzing runs. %

\textit{Weird Peer.}
We use our LLVM pass and its runtime with the abovementioned changes to instrument the weird peer. In addition, we implement logic to detect when the server is ready for client connections. For this, we add support for hooking the functions \verb|listen|, \verb|bind|, and \verb|connect|. Hooking these functions allows us to determine when a server is ready to accept connections and whether a client tried to connect to the server. The latter becomes handy if a mutation causes a weird client to crash before being able to connect to the server.  If a mutation leads to a client crash before connecting, we discard that fuzzing run, as the crashed client cannot impact the server's operation.

\textit{Target Peer.}
For the target peer, we leverage AFL++ in version v4.08 for its LLVM-based coverage instrumentation. Beyond that, we use the same hooking mechanism, modified LLVM pass and runtime as for the weird peer. This is necessary as the server can be either the weird peer or the target peer. %

%% file: content/05_evaluation.tex
\section{Evaluation}\label{sec:evaluation}

\begin{figure*}[ht]
    \centering
    \graphicspath{{eval-plots/}}
    \def\svgwidth{\linewidth}
    \begin{footnotesize}
        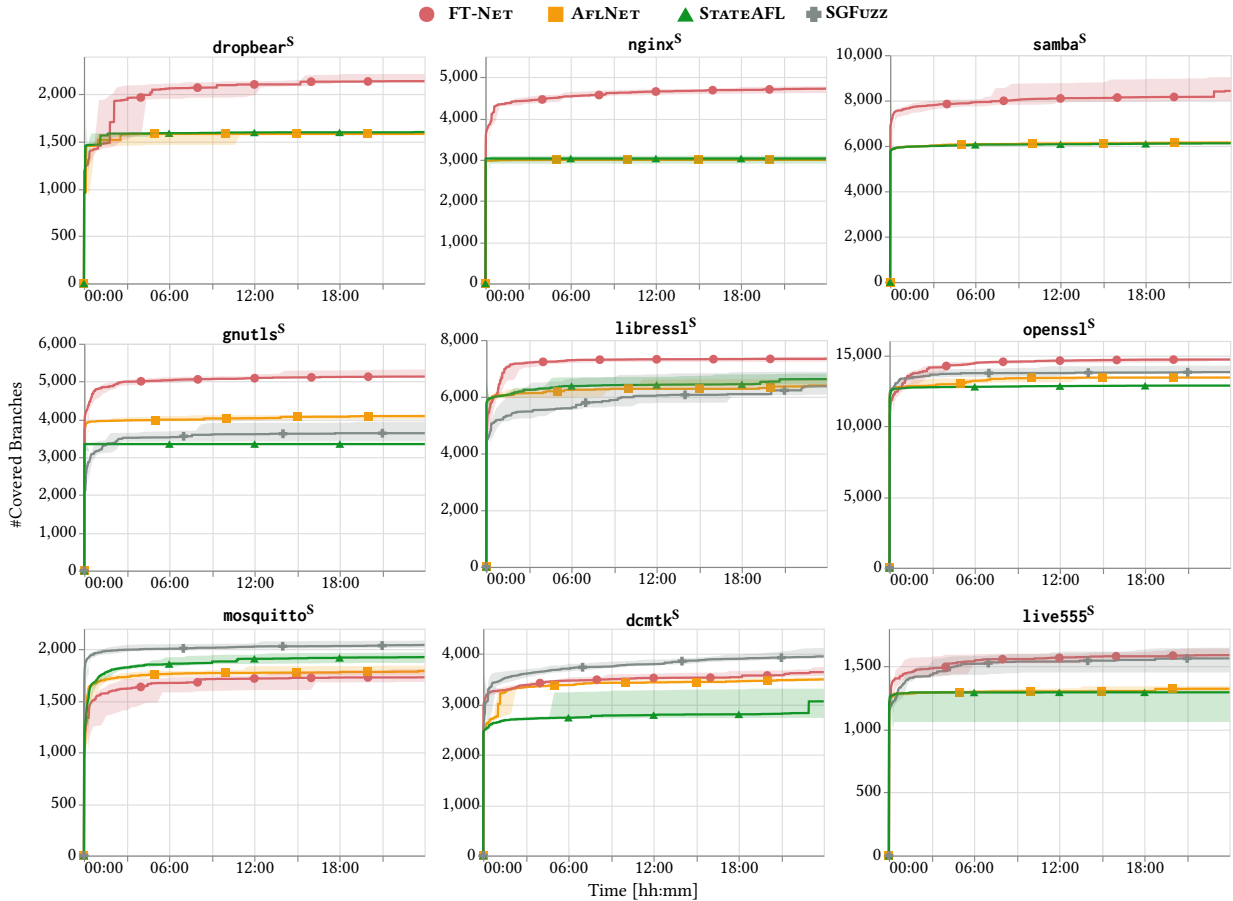
    \end{footnotesize}
    \vspace{-3em}
    \caption{The coverage (in \llvmcov branches) various fuzzers achieve over ten 24h runs on different targets. We display the median run and 66\% intervals. }%
    \label{fig:eval:coverage_comp}
\end{figure*}

We evaluate how our implementation \fuzztructionnet performs compared to other network application fuzzers and whether it is capable of uncovering new bugs in real-world targets. To this end, we conduct three different experiments.
\begin{inparaenum}
\item We benchmark our approach regarding code coverage and compare it to state-of-the-art network fuzzers.
\item Based on the coverage benchmark data, we evaluate the bug-finding capabilities of each individual fuzzer by analyzing all bugs that have been found during the coverage experiment.
\item Then we study the capabilities of \fuzztructionnet by testing additional widely deployed software projects, including clients that no other network fuzzer is designed to fuzz. %
\end{inparaenum}

\subsection{Setup}
Before discussing our experiments and their outcome in-depth, we outline our setup and present targets we use throughout our evaluation. Overall, we follow the recommendations by Klees~\etal~\cite{klees2018evaluating}. %

\textbf{Hardware Environment.}
To ensure fair experiments, we used the same hardware configuration for all our experiments: an Intel Xeon Gold 5320 CPU @ 2.20GHz (52 physical cores), 256 GB of RAM, and SSD memory as backing storage. To model realistic conditions, we spawned 13 fuzzer worker processes per fuzzer and pinned these workers to 13 (physical) cores. This allowed us to perform four experiments in parallel.

\textbf{Fuzzers.}
We compare \fuzztructionnet against three state-of-the-art fuzzers: \aflnet~\cite{pham2020aflnet}, \sgfuzz~\cite{ba2022sgfuzz}, and \stateafl~\cite{natella2022stateafl}. 
We considered evaluating against other existing works like \bleem~\cite{luo2023bleem}, but their code is not publicly available, and the authors could not share the code when we contacted them. Other protocol fuzzers, such as \boofuzz~\cite{boofuzz} or \peach~\cite{peach}, are available, but they offer only a limited range of protocol specifications by default, which is insufficient for testing more complex or diverse targets. Additionally, related work has found these tools to perform worse compared to fuzzers we included on our evaluation targets~\cite{luo2023bleem,pham2020aflnet}. In summary, we configured the tested fuzzers as follows:

\begin{enumerate}
    \item \textbf{\aflnet~\cite{pham2020aflnet}} (\texttt{62d63a5}): We enabled its state-aware mode (\texttt{-E}) and the region-level mutation operators (\texttt{-R}) by default. In addition, we added a \texttt{NOP} protocol that does not provide any state feedback during fuzzing but allows for testing targets that are \emph{not} supported by \aflnet. We used this protocol for the targets \mosquittoServer, \nginxServer, and \sambaServer. %
    \item \textbf{\sgfuzz~\cite{ba2022sgfuzz}} (\texttt{00dbbd7}): To use the distinct values of \texttt{enum} variables, \sgfuzz needs to preprocess the target's source code with a Python script that instruments assignments of \texttt{enum} variables. Unfortunately, this script crashes for some source files, aborting instrumentation. We identified and removed offending files from preprocessing.
    \item \textbf{\stateafl~\cite{natella2022stateafl}} (\texttt{d923e22}): As this tool is based on \aflnet (but uses memory allocations to deduce protocol state instead of handcrafted protocol parsers), we used it with the same flags as \aflnet. \stateafl's instrumentation sometimes misses \texttt{free} operations, compromising their memory allocation tracking. As the fuzzer continues to access these already freed allocations to determine the protocol state, segmentation faults often occur (resulting in false positive crashes). Additionally, \stateafl does not support certain functions, such as vectorized read and write operations, and initially refused to start if no network-related operations were detected, such as data written or read from a socket. To address these issues, we added support for multiple \texttt{libc} functions, including \texttt{readv} and \texttt{writev}. Despite these improvements, we encountered persistent issues with memory tracking for some targets, thereby limiting the overall robustness of \stateafl on these targets.
\end{enumerate}

\textbf{Target Applications.}
For the comparative evaluation and the real-world bug-finding experiment, we selected \numtargets targets from \numuniquetargets software projects in total. We also selected a second application acting as weird peer for each target. In all cases, we used a test client and server shipped by the software project, such that no manual work was needed to derive the weird peer. %
In the following, we indicate whether we refer to the client or server of a project by using superscripts, \ie \gnutlsClient and \gnutlsServer refer to the client and server of the \gnutls software suite, respectively. 

When selecting suitable targets, we aimed to compile a set of applications that represent realistic targets for fuzzing network applications. %
While recommended, benchmarks~\cite{natella2021profuzzbench} do not support our non-traditional way of fuzzing, making it impossible for us to use them.
We selected targets according to the following criteria:
\begin{enumerate}
    \item Comparability with related work: The set of evaluated applications should intersect with the ones used by other network application fuzzers. 
    \item Active projects: The selected targets should be actively maintained and relevant in practice.
    \item Optimally, the targets are thoroughly tested and have been fuzzed before. This allows us to quantify our new approach's impact rather than attributing findings to the fact that a project is tested for the first time.
\end{enumerate}

To meet the first criterion, we selected a subset of the applications tested as part of the evaluation of the other fuzzers. From the evaluations of \aflnet and \sgfuzz, we selected \livefivefivefiveServer, \dcmtkServer, and \opensslServer. %
Since both fuzzers found over 90\% of their bugs in \livefivefivefiveServer and \dcmtkServer, we expected these targets to be excellent candidates for our evaluation. %

To ensure we meet criteria (2) and (3), we selected the remaining targets based on whether they are shipped with fuzzing harnesses or are part of OSS-Fuzz~\cite{ossfuzz} and if they seem to be actively maintained and do not have stale bugs in their respective bug trackers.
Table~\ref{tab:eval:fuzz_support} lists all targets we selected, outlining whether these targets have fuzzing harnesses as part of their repository and if they are integrated into \ossfuzz. Out of the \numuniquetargets applications, only two are not integrated into \ossfuzz, \dcmtk and  \livefivefivefive. Interestingly, both have been the primary source of new bugs found by \aflnet and \sgfuzz. Table~\ref{tab:appendix:target_configuration} in the Appendix lists the exact configuration of each target.

\begin{table}[t]
    \centering
    \caption{Information on targets that are part of our evaluation. A complexity (Cmpl.) of \texttt{easy} indicates the protocol is amenable to byte-oriented mutations, while we use \texttt{hard} to indicate targets using checks and/or encryption to protect communication. We also survey whether the targets offer fuzzing harnesses (H) or are even included in \ossfuzz' testing (OF). The applications in the \generatorsc column are the ones used by \fuzztructionnet as a peer for the target. The targets above the line were selected for our comparative evaluation.}
    \label{tab:eval:fuzz_support}
    \begin{adjustbox}{max width=\columnwidth}
    \begin{tabular}{lllllcc}
    \toprule
                             &&                     && \multicolumn{3}{c}{\textbf{Fuzzing Support}}                  \\
        \multirow{-2}{*}{\textbf{Target}}     &\multirow{-2}{*}{\textbf{Protocol}}& \multirow{-2}{*}{\textbf{Version}}    &\multirow{-2}{*}{\textbf{\generatorsc}}& \textbf{Cmpl.} & \textbf{H} & \textbf{OF}    \\
    \midrule %
        \dropbearServer                 &SSH& 9925b00& \dropbearClient & \texttt{hard}         & \cmark                        & \cmark \\
        \dcmtkServer                    &DICOM& 1549d8c& \dcmtkClient & \texttt{easy}            & \xmark                        & \xmark \\
        \gnutlsServer                   &TLS& e840a07& \gnutlsClient & \texttt{hard}         & \cmark                        & \cmark \\
        \libresslServer                 &TLS& fbb21ed& \libresslClient & \texttt{hard}         & \cmark                        & \cmark \\
        \livefivefivefiveServer         &RTP& 2023.06.14& \livefivefivefiveClient & \texttt{med.}            & \xmark                        & \xmark \\
        \mosquittoServer                &MQTT& 3923526& \mosquittoClient & \texttt{easy}              & \xmark                        & \cmark \\
        \nginxServer                    &HTTP/3& 6b1bb99& \ngtcptwoClient & \texttt{hard}         & \cmark                        & \cmark \\
        \opensslServer                  &TLS& 7b649c7& \opensshClient & \texttt{hard}         & \cmark                        & \cmark \\
        \sambaServer                    &SMB& 95474d8& \sambaClient & \texttt{hard}         & \cmark                        & \cmark \\
    \midrule %
    \apacheServer                 &HTTP/2& a751ae5& \curlClient & \texttt{hard}         & \cmark                        & \cmark \\
    \pjsipServer                    &SIP& 12d0468& \pjsipClient & \texttt{easy}              & \cmark                        & \cmark \\
    \curlClient                 &HTTP/3& de7b3e8 & \nginxServer & \texttt{hard}         & \cmark                        & \cmark \\
    \dropbearClient                 &SSH& 9925b00& \dropbearServer & \texttt{hard}         & \cmark                        & \cmark \\
    \libresslClient                 &TLS& fbb21ed& \libresslServer & \texttt{hard}         & \cmark                        & \cmark \\
    \ngtcptwoServer                 &HTTP/3& f3f15b6 & \ngtcptwoClient & \texttt{hard}         & \cmark                        & \cmark \\
    \opensshClient                   & SSH& 86bdd38& \dropbearServer & \texttt{hard}& \cmark                        &\cmark \\
    \bottomrule
    \end{tabular}
    \end{adjustbox}
\end{table}

\textbf{Target Configuration.}
We compiled all targets using the compiler pass provided by the respective fuzzer. Since most targets have already been extensively tested, we deployed ASAN for each fuzzer on all fuzzing targets to detect bugs that generally would not cause a crash. Where applicable, we disabled custom allocators to avoid bugs being shadowed by ASAN's inability to determine when a memory chunk was freed.

Since all other network application fuzzers rely on replay-based techniques, \ie traffic of a real client/server communication must be recorded and used as seed, their fuzzing efforts are impeded by sources of randomness such as session identifiers or ephemeral tokens. Notably, this limitation is shared by all publicly available network fuzzers and is not a result of our fuzzer selection. To address this shortcoming, we attempted to disable sources of randomness in all fuzzing targets. This includes the use of static session IDs as well as seeding random number generators with constant seeds. Furthermore, we removed functions typically used to reseed the PRNGs during runtime. However, it is important to note that it is impossible to encompass \emph{all} sources of randomness simply due to the complexity of the tested targets. Furthermore, we only patched the target applications themselves and the libraries shipped alongside them. We did not patch libraries that are pulled from the system as external dependencies, since this would have been infeasible to accomplish. 
We provide the patches as part of our artifact at~\url{https://github.com/fuzztruction/fuzztruction-net-experiments}. We stress that patching out randomness is required by the other tested tools, not \fuzztructionnet.

For \fuzztructionnet, we compiled the targets using a custom version of \aflpp v4.08c, as detailed in Section~\ref{sec:implementation}. Furthermore, we enabled laf-intel~\cite{lafintel} via setting \verb|AFL_LLVM_LAF_ALL|. We compiled the \generator applications using our compiler pass, allowing us to mutate the application during execution dynamically. Note that we disabled randomness as for the other approaches to ensure a fair evaluation, even though our approach does not necessarily require this step.

\textbf{Seeds.}
Since \fuzztructionnet uses a mutated client or server application to produce fuzzing inputs for the SUT, it does not require seed files but another peer. For the other fuzzers, we prepared seeds by recording the network traffic between each client/server pair (same pair as used by \fuzztructionnet). Thus, each approach had the same initial conditions for its fuzzing campaign.

\textbf{Coverage.}
To calculate coverage, we used a version of the target compiled with LLVM's Source-based Code Coverage, which enabled us to measure the coverage for each input processed by the target being tested. LLVM-Cov specifically uses the branches in the source code as the units for coverage measurement. This approach ensures that the coverage results are independent of the machine code produced by the compiler. As a result, these findings are consistent across various compiler versions and even among binaries that have undergone different instrumentation processes.

\subsection{Coverage Experiments}%
\label{sec:coverage_experiments}
While fuzzing primarily aims to identify bugs, coverage is a well-established and widely used proxy metric for comparing the efficiency and effectiveness of different fuzzers~\cite{boehme2022_covreliability}. 
For our evaluation, we selected nine different targets (\cf first half of Table~\ref{tab:eval:fuzz_support}) and fuzzed them for 24 hours. %
Given the stochastic nature of fuzzing, we repeated each experiment ten times to ensure reliability~\cite{klees2018evaluating}.

We encountered challenges in deploying \sgfuzz on three targets: \dropbearServer, \sambaServer, and \nginxServer. For \nginxServer, the primary issue was the lack of UDP support in \sgfuzz. The \sambaServer target refused to compile when instrumented with the \sgfuzz instrumentation. The reason for \sgfuzz failing on \dropbearServer was the lack of support for targets using \texttt{execve}, which is mandatory if \dropbearServer is not intended to be used in one-shot mode (\ie it terminates after processing one connection), which is also not support by \sgfuzz.  

The results of the coverage experiment are shown in Figure~\ref{fig:eval:coverage_comp}. Overall, our prototype \fuzztructionnet outperforms all other fuzzers in six out of nine cases. In one of the nine cases, namely for the target \livefivefivefiveServer, our performance is on par with \sgfuzz. However, \fuzztructionnet underperforms \sgfuzz for \dcmtkServer, and both \sgfuzz and \stateafl showed better performance for the \mosquittoServer target.

\begin{table}[t]
    \centering
    \caption{Statistical evaluation of coverage results. We use a bootstrap-based algorithm~\cite{schloegel2024sok} to determine significance and Vargha and Delaney's $\hat{A}_{12}$ test to measure effect size~\cite{vargha2000critique}. If the difference is significant, we print the effect size in bold.}
    \begin{adjustbox}{max width=\columnwidth}
    \begin{tabular}{llcr}
        \toprule
        \multicolumn{1}{c}{\textbf{Target}} & \textbf{Best Competitor} & \textbf{$\hat{A}_{12}$ effect size} \\
        \midrule
\dropbearServer         & \stateafl   & \textbf{+L(1.00)} \\
\dcmtkServer            & \sgfuzz     & \textbf{-L(0.00)} \\
\gnutlsServer           & \aflnet     & \textbf{+L(1.00)} \\
\libresslServer         & \stateafl   & \textbf{+L(0.94)} \\
\livefivefivefiveServer & \sgfuzz     &         +S(0.63)  \\
\mosquittoServer        & \sgfuzz     & \textbf{-L(0.00)} \\
\nginxServer            & \stateafl   & \textbf{+L(1.00)} \\
\opensslServer          & \sgfuzz     & \textbf{+L(1.00)} \\
\sambaServer            & \aflnet     & \textbf{+L(1.00)} \\
         \bottomrule
    \end{tabular}
    \end{adjustbox}
    \label{tab:eval:statistics}
\end{table}

\textbf{Statistical evaluation.}
We statistically evaluate our results in line with recommended best practices~\cite{klees2018evaluating}. To this end, we use a bootstrap-based two-sample t-test~\cite{schloegel2024sok} and measure effect size via Vargha's and Delaney's $\hat{A}_{12}$ test~\cite{vargha2000critique}, comparing \fuzztructionnet to the best other tool (based on the coverage of the median run). As shown in Table~\ref{tab:eval:statistics}, the effect size is large (\textgreater 0.71~\cite{vargha2000critique}) in all cases with the exception of \livefivefivefiveServer, where we observe no statistically significant difference. In two cases, \dcmtkServer and \mosquittoServer, we see a negative effect size, as \fuzztructionnet performed worse than \sgfuzz. Overall, the statistical evaluation confirms our intuition that \fuzztructionnet is better on six out of nine targets, worse on two, and tied with \sgfuzz on \livefivefivefiveServer. 

Next, we analyze these results in more detail, first focusing on cases where \fuzztructionnet succeeded before then taking a closer look at the targets where \fuzztructionnet was not the fuzzer with the best performance. 
We reference the challenges outlined in Section~\ref{sec:challenges}.

\paragraph{Targets with good performance.}
For six targets, \fuzztructionnet shows excellent performance characteristics, outperforming the best competitor by at least 6\% (on \opensslServer) but up to 56\% (\nginxServer), with an average improvement of 16\% over all nine targets.
When analyzing the coverage achieved by various fuzzers, it becomes clear that other fuzzers encounter significant obstacles with certain targets such as \sambaServer, \nginxServer, and \dropbearServer. The reason is that these targets use ephemeral values (challenge C1), such as session identifiers, or encryption (challenge C2). %

For the TLS libraries \gnutlsServer, \libresslServer, and \opensslServer, the other fuzzers successfully replayed the initial seed provided by us. This was only possible because we modified these targets' Pseudo Random Number Generators (PRNGs) to yield predictable outputs. However, the fuzzers still failed to initiate any new successful TLS handshakes beyond using the initial seed file. This failure occurs because any slight alteration in the seed file requires a recalculation of the message authentication codes used during the handshake (C1, C2), due to the protocol's inherent design.
The situation is even more complicated for \sgfuzz because of its in-process approach. Although the PRNGs produce predictable values, they are not reset by \sgfuzz after processing a connection. Therefore, in subsequent fuzzing iterations, the random numbers used during the TLS handshake differ from those in the seed file (C1), leading to earlier rejection of generated messages. This limitation impedes \sgfuzz's ability to explore as much error-handling code as the other fuzzers.

\fuzztructionnet, in contrast, does not suffer from any of these limitations, leading to the observed outcome. Upon closer inspection, we found that some of the other fuzzers covered more error-handling code than \fuzztructionnet, suggesting that using multiple fuzzers may yield a beneficial synergy.

\paragraph{Targets with bad performance.}
In the following, we analyze in detail the targets where \fuzztructionnet underperformed our expectations.

\begin{figure}[t]
    \centering
    \graphicspath{{eval-plots/}}
    \def\svgwidth{\linewidth}
    \begin{footnotesize}
        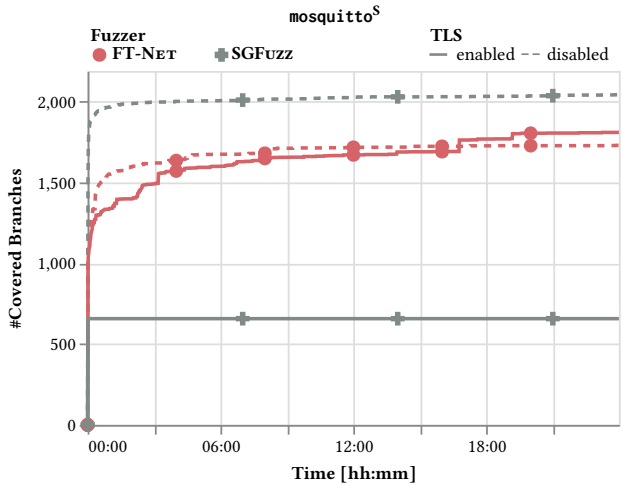
    \end{footnotesize}
    \vspace{-2em}
    \caption{Coverage on \mosquittoServer when enabling TLS support as typical for real-world use. We compute coverage computed only for the target, not the TLS library implementation. TLS stalls \sgfuzz but not \fuzztructionnet. 
    }%
    \label{fig:eval:mosquito_tls}
\end{figure}

\textbf{\mosquittoServer.}
On this target, \fuzztructionnet falls short compared to all other tools. When investigating this behavior, we found that the characteristics of the MQTT protocol are the root cause: MQTT handles a very limited set of message types, primarily \texttt{subscribe} and \texttt{unsubscribe} messages. The MQTT protocol is straightforward and lightweight to be used on embedded devices, making it relatively easy for fuzzers to generate syntactically correct messages purely by chance.
We hypothesize that it is, hence, less important to generate ``good'' inputs by faulting the peer than to generate as many (random) inputs as quickly as possible. 
Further, our data shows that more than 55\% of the inputs crafted by the fuzzers and sent to the \mosquittoServer were shorter than 32 bytes, indicating the simplicity of the protocol format. Such compact input sequences are ideal targets for bit-level mutational fuzzers, which excel in enumerating through small variations of input data. If MQTT used more complex mechanisms, such as session IDs that require maintaining state across messages, this might have significantly hindered the effectiveness of such basic fuzzing techniques. Another interesting effect that we observed is that \sgfuzz was the only fuzzer to cover code related to data persistence. The reason for that is that \sgfuzz is the only tool using an in-memory fuzzing approach, which causes \mosquittoServer to persist the server data after a number of connections have been processed. This does not happen for the other fuzzers, which terminate the target after each iteration. 

However, depending on which features are enabled, our approach can be beneficial even for targets with such a simple protocol specification. 
To showcase this claim, we enabled TLS support for \mosquittoServer, which is arguably a realistic scenario, and ran the fuzzers again. As shown in Figure~\ref{fig:eval:mosquito_tls} (with full data for all fuzzers in Figure~\ref{fig:eval:mosquitto_tls_diff_all} in the Appendix), traditional bit-level mutation fuzzers then struggle with this target. 
The coverage (measured \emph{only} for the target, excluding the TLS library implementation itself) for \fuzztructionnet remains unimpeded, while other fuzzers are less effective and can no longer fuzz the target thoroughly.

\begin{figure}[t]
    \centering
    \graphicspath{{eval-plots/}}
    \def\svgwidth{\linewidth}
    \begin{footnotesize}
        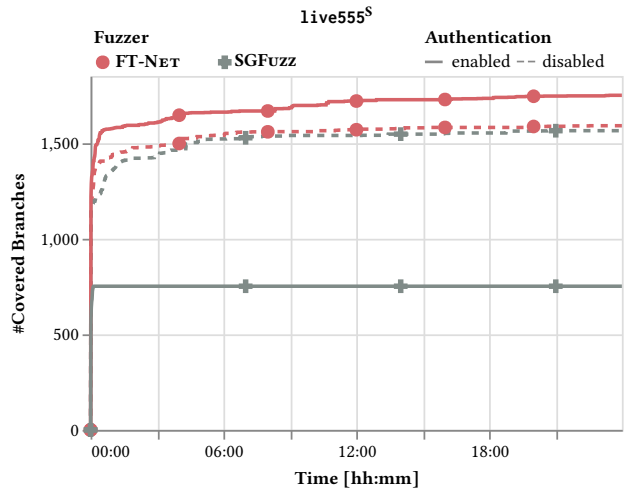
    \end{footnotesize}
    \vspace{-2em}
    \caption{Coverage on \livefivefivefiveServer with HTTP Digest Access Authentication~\cite{rfc7616} enabled. We still provide replay-based fuzzers with a recording of a successful authentication (HTTP plain text). }%
    \label{fig:eval:live555_auth}
\end{figure}

\textbf{\dcmtkServer.}
For \dcmtkServer, \sgfuzz outperforms \fuzztructionnet. Similar to the situation with \mosquittoServer, the subset of the DCMTK protocol used by the target lacks complex primitives that challenge traditional fuzzers. The main function of \dcmtkServer involves receiving an x-ray image along with accompanying metadata sent by the client.
Analyzing the coverage differences of \fuzztructionnet and \sgfuzz, we found that the coverage advantage primarily stems from parsing functions that, for example, convert integer or float values into strings or vice versa.
For fuzzing these kinds of functions, approaches with a high throughput rate, like \sgfuzz, are inherently better suited and, therefore, have an advantage.

\textbf{\livefivefivefiveServer.}
The third case where \fuzztructionnet did not perform better than all state-of-the-art tools, but was on par with \sgfuzz, is \livefivefivefiveServer. Notably, we patched the target (as recommended by the authors of \sgfuzz~\cite{ba2022sgfuzz}) such that it uses a constant value as session IDs, such that replayed messages are still accepted by the server. In terms of source code coverage, both fuzzers covered mostly the same code. 
However, one notable area that neither fuzzer addressed is the server’s authentication functionality. This feature was intentionally disabled for our experiment. In a real-world application, similar to TLS in the \mosquittoServer scenario, authentication is likely enabled. The authentication system used by \livefivefivefiveServer incorporates a server-selected nonce as part of HTTP authentication. Despite the simplicity, this renders traditional fuzzers ineffective as evident from Figure~\ref{fig:eval:live555_auth} (data for all fuzzers in Figure~\ref{fig:eval:live555_auth_diff_all} in the Appendix). With this authentication enabled, the other fuzzers did not manage to establish a single successful authentication, in contrast to \fuzztructionnet, even though the provided seed contained the exact HTTP plain text packets recorded during a successful authentication.

\subsection{Finding New Vulnerabilities}%
To measure the bug-finding capability of our approach, we conducted two experiments. First, we compared the number of unique bugs found by all fuzzers based on the deduplicated crashes found during the coverage evaluation. In the second experiment, we tested \fuzztructionnet's ability to find bugs in an additional set of targets, outlined in the bottom half of Table~\ref{tab:eval:fuzz_support}. In particular, these targets include several clients that no other network application fuzzer can test.
All selected targets were evaluated using the most recent versions available at the time of writing. Interestingly, most of the targets are integrated into OSS-Fuzz and provide, in addition, fuzzing harnesses for a variety of their interfaces. Therefore, it can be assumed that most bugs found during our evaluation eluded previous fuzzing efforts. 

To map found crashes to underlying bugs, we used a two-staged schema for pre-filtering:
\begin{inparaenum}
    \item Crashes were bucketed according to the last five called functions of the ASAN stack traces that are generated when a crash is found.
    \item The resulting buckets were manually triaged and mapped to unique bugs.
\end{inparaenum}

\begin{table}[t]
    \centering
    \caption{Unique bugs found by the different fuzzers. We derive this number by first bucketing found crashes via the last five functions on the call stack in the ASAN report and then deduplicating them manually. %
    }%
    \label{tab:eval:bugs_comp}
    \begin{adjustbox}{max width=\columnwidth}
    \begin{tabular}{lS[table-format=2]S[table-format=1]S[table-format=1]S[table-format=1]r}
        \toprule
        \multicolumn{1}{c}{\textbf{Target}} & \fuzztructionnet & \aflnet & \sgfuzz & \stateafl & \#Unique Bugs \\
        \midrule
\dropbearServer         & \color{black!40}0 & \color{black!40}0 & \color{black!40}0 & \color{black!40}0 & \color{black!40}0 \\
\dcmtkServer            & 2 & 2 & 2 & 1 & 2 \\
\gnutlsServer           & 2 & \color{black!40}0 & \color{black!40}0 & \color{black!40}0 & 2 \\
\libresslServer         & \color{black!40}0 & \color{black!40}0 & \color{black!40}0 & \color{black!40}0 & \color{black!40}0 \\
\livefivefivefiveServer & 2 & \color{black!40}0 & 1 & \color{black!40}0 & 3 \\
\mosquittoServer        & \color{black!40}0 & \color{black!40}0 & \color{black!40}0 & \color{black!40}0 & \color{black!40}0 \\
\nginxServer            & 3 & \color{black!40}0 & \color{black!40}0 & \color{black!40}0 & 3 \\
\opensslServer          & \color{black!40}0 & \color{black!40}0 & \color{black!40}0 & \color{black!40}0 & \color{black!40}0 \\
\sambaServer            & \color{black!40}0 & \color{black!40}0 & \color{black!40}0 & \color{black!40}0 & \color{black!40}0 \\
    \midrule
    Sum                 & 9 & 2 & 3 & 1 & 10 \\
    \bottomrule
    \end{tabular}
    \end{adjustbox}
\end{table}

\paragraph{Comparative Evaluation.}%
\label{eval:sec:bug_finding_comperative}
We present the results of our comparison with other fuzzers in Table~\ref{tab:eval:bugs_comp}.
A total of 10 unique bugs were identified across all targets. Out of these, \fuzztructionnet successfully detected 9 bugs, missing only a single bug in \livefivefivefiveServer. In general, the effectiveness of each fuzzer in identifying bugs aligns with its coverage performance. Targets where \fuzztructionnet excelled in coverage also yielded a higher number of detected bugs compared to other fuzzers. The only exception is \mosquittoServer, where \sgfuzz achieved significantly higher coverage than \fuzztructionnet, yet none of the fuzzers found any bugs. \stateafl, on the other hand, identified only a single bug. This limited detection capability can be attributed to frequent crashes caused by its memory allocation tracking instrumentation, which often prevented the fuzzer from making meaningful progress.

\begin{table*}[t]
    \centering
    \caption{Overview of the \numbugs bugs we found in different targets. All bugs have been responsible disclosed. \textbf{Weird Peer} indicates the program in which we induced faults to generate the input uncovering the bug. The \textbf{Status} column links to the CVE or bug reports, unless a bug was reported via email. All CVEs have been fixed.} %
    \label{tab:Bugs}
    \begin{adjustbox}{max width=\linewidth}
 \begin{tabular}{lllll}
        \toprule
        \textbf{Target} & \textbf{Weird Pe\rlap{er}} & \textbf{Status} & \textbf{Type}&\textbf{Description}\\
        \midrule
        \nginxServer    & \ngtcptwoClient & \textit{fixed}     & Heap OOB read  & \verb|ngx_quic_parse_transport_param| uses user-controlled length value   \\ %
        \nginxServer    & \ngtcptwoClient & \bugreportcve{CVE-2024-32760}     & Heap OOB writ\rlap{e}  & Duplication of an entry in the HTTP3 dynamic table, while the table is too small \\ %
        \nginxServer    & \ngtcptwoClient & \bugreportcve{CVE-2024-31079}     & Heap UAF  &  During an infinite recursion, the recurringly accessed memory pool is freed  \\ %
        \nginxServer    & \ngtcptwoClient & \bugreportcve{CVE-2024-3416}     & Info. leak. & QUIC packets can cause worker processes to leak previously freed memory \\ %
        \nginxServer    & \ngtcptwoClient & \bugreportcve{CVE-2024-35200}     & Nullptr deref & Access to a variable  holding the user-agent header value is unexpectedly null  \\ %
        \curlClient     & \nginxServer    & \textit{fixed} & Nullptr deref  & \verb|ngtcp2_ksl_begin| in the \verb|ngtcp2| library tries to accesses the head of a list that is null \\
        \curlClient     & \nginxServer    & \textit{fixed} & Assertion  & In the \verb|nghttp3| library, an assertion checking if enough data is remaining is triggered \\
        \apacheServer   & \curlClient & \textit{reported}     & Heap UAF  & \verb|apr_file_write| attempts to write error log after freeing its path  \\ %
        \apacheServer   & \curlClient & \textit{reported}     & Assertion   & While adding a key-value pair into an APR table, an assertion is triggered \\
        \opensshClient  & \dropbearServer & \textit{fixed} & Heap UAF  & The asynchronous callback \verb|verify_host_key| accesses the previously freed host key \\
        \dropbearClient & \dropbearServer & \blackhref{https://github.com/mkj/dropbear/issues/285}{\textit{fixed} (\#285)} & Assertion  & \verb|signkey_type_from_signature| tries to cast an invalid integer to an \texttt{enum} \\
        \gnutlsServer   & \gnutlsClient   & \blackhref{https://gitlab.com/gnutls/gnutls/-/issues/1529}{\textit{fixed} (\#1529)}     & Nullptr deref  & During a retry-handshake, \verb|_gnutls_cipher_auth| dereferences a null pointer  \\
        \gnutlsServer   & \gnutlsClient   & \blackhref{https://gitlab.com/gnutls/gnutls/-/issues/1534}{\textit{fixed} (\#1534)} & Nullptr deref   &  \verb|_gnutls_figure_common_ciphersuite| dereferences a null ptr if a PSK cipher is used \\ %
        \libresslClient & \libresslServer & \blackhref{https://github.com/libressl/portable/issues/1037}{\textit{fixed} (\#1037)} & Nullptr deref   & Attempting to print key material in \verb|ssl_print_tmp_key| after it has been freed \\
        \multirow{2}{*}{\ngtcptwoServer} & \multirow{2}{*}{\ngtcptwoClient} & \multirow{2}{*}{\blackhref{https://gitlab.com/gnutls/gnutls/-/issues/1529}{\textit{fixed} (\#1529)}}     & \multirow{2}{*}{Assertion}  & The \verb|wolfssl| library returns a TLS session using the unauthenticated \verb|CTR| AES mode, \\
        & & & & even though the \verb|CCM| mode was negotiated  \\ %
        \ngtcptwoServer & \ngtcptwoClient &  \blackhref{https://github.com/wolfSSL/wolfssl/issues/7406}{\textit{fixed} (\#7406)}   & Heap OOB writ\rlap{e}  & In \verb|wolfSSL_read_early_data|, header bytes are written to a insufficiently sized buffer \\
        \pjsipServer    & \pjsipClient    & \textit{fixed} & Heap OOB read  & A timer object is deleted before it expires, causing an out-of-bound read   \\
        \pjsipServer    & \pjsipClient    & \textit{fixed} & FP Exception  & Proposing a clock rate of 0 for an audio stream causes a division by zero \\
        \pjsipServer    & \pjsipClient    & \textit{fixed} & Heap OOB writ\rlap{e} & Dumped source address of an SDP message is copied into too small buffer  \\ %
        \dcmtkServer    & \dcmtkClient    & \bugreportcve{CVE-2024-34508} & Nullptr deref  & \verb|checkAndProcessSTORERequest| accesses the value of an element pointing to null  \\
        \dcmtkServer    & \dcmtkClient    & \bugreportcve{CVE-2024-34509} & Nullptr deref  & \verb|DIMSE_parseCmdObject| attempts to access a string that points to null \\
        \livefivefivefiveServer     & \livefivefivefiveClient    & \textit{reported} & Heap UAF  & \verb|sendDataOverTCP| sends MP3 bytes after freeing the \verb|RTSPClientConnection|  \\ %
        \livefivefivefiveServer     & \livefivefivefiveClient    & \textit{reported} & Heap UAF & During execution of \verb|handleHTTPCmd_TunnelingPOST|, the processed input is freed \\ %
        \bottomrule
    \end{tabular}
    \end{adjustbox}
    \begin{flushleft}
    \end{flushleft}
\end{table*}

\paragraph{Real-World Applicability.} %
\label{eval:sec:bug_finding_real_world}
To test the real-world applicability of \fuzztructionnet, we fuzzed further targets, including many client applications that other network fuzzers cannot test. A list of reported bugs found by \fuzztructionnet is shown in Table~\ref{tab:Bugs}. Overall, \fuzztructionnet found bugs in a variety of security-relevant targets, including the OpenSSH client and the webservers HTTPd and Nginx (where two CVEs have been assigned). The detected bugs cover the full spectrum of typical memory safety violations, ranging from nullptr dereferences to out-of-bounds memory accesses on the heap to use-after-frees.

\paragraph{Case Study: Nginx.}
One of the bugs \fuzztructionnet found in Nginx allows for an 8-byte out-of-bounds write in the HTTP/3 dynamic table, a critical component used within the QPACK header compression mechanism of QUIC. The HTTP/3 dynamic table stores previously sent headers to optimize header compression over QUIC, thereby reducing redundancy and enhancing the efficiency of subsequent requests. \fuzztructionnet triggered the bug via two steps: First, the weird peer \ngtcptwoClient adjusted the dynamic table size to a new capacity by sending a ``Set Dynamic Table Capacity'' message. Then, via a ``Duplicate'' message, it requested to duplicate an entry in the dynamic table. Crucially, the size requested initially could not accommodate the duplication, leading to an overflow. This overflow introduced the potential for memory corruption, posing a serious security risk in the management of HTTP/3 headers in Nginx. It is noteworthy that none of the other fuzzers found this bug, as their design did not allow them to access the key generated during the TLS handshake.

%% file: eval-plots/group_1.pdf_tex
\begingroup%
  \makeatletter%
  \providecommand\color[2][]{%
    \errmessage{(Inkscape) Color is used for the text in Inkscape, but the package 'color.sty' is not loaded}%
    \renewcommand\color[2][]{}%
  }%
  \providecommand\transparent[1]{%
    \errmessage{(Inkscape) Transparency is used (non-zero) for the text in Inkscape, but the package 'transparent.sty' is not loaded}%
    \renewcommand\transparent[1]{}%
  }%
  \providecommand\rotatebox[2]{#2}%
  \newcommand*\fsize{\dimexpr\f@size pt\relax}%
  \newcommand*\lineheight[1]{\fontsize{\fsize}{#1\fsize}\selectfont}%
  \ifx\svgwidth\undefined%
    \setlength{\unitlength}{884.25bp}%
    \ifx\svgscale\undefined%
      \relax%
    \else%
      \setlength{\unitlength}{\unitlength * \real{\svgscale}}%
    \fi%
  \else%
    \setlength{\unitlength}{\svgwidth}%
  \fi%
  \global\let\svgwidth\undefined%
  \global\let\svgscale\undefined%
  \makeatother%
  \begin{picture}(1,0.7336726)%
    \lineheight{1}%
    \setlength\tabcolsep{0pt}%
    \put(0.05133698,0.37645576){\color[rgb]{0,0,0}\rotatebox{90}{\makebox(0,0)[t]{\lineheight{1.25}\smash{\begin{tabular}[t]{c}\#Covered Branches\end{tabular}}}}}%
    \put(0.51603034,0.04677569){\color[rgb]{0,0,0}\makebox(0,0)[t]{\lineheight{1.25}\smash{\begin{tabular}[t]{c}Time [hh:mm]\end{tabular}}}}%
    \put(0,0){\includegraphics[width=\unitlength,page=1]{group_1.pdf}}%
    \put(0.3946787,0.70197983){\color[rgb]{0,0,0}\makebox(0,0)[t]{\lineheight{1.25}\smash{\begin{tabular}[t]{c}\textbf{\toolname}\end{tabular}}}}%
    \put(0,0){\includegraphics[width=\unitlength,page=2]{group_1.pdf}}%
    \put(0.48770097,0.70197983){\color[rgb]{0,0,0}\makebox(0,0)[t]{\lineheight{1.25}\smash{\begin{tabular}[t]{c}\textbf{\aflnet}\end{tabular}}}}%
    \put(0,0){\includegraphics[width=\unitlength,page=3]{group_1.pdf}}%
    \put(0.59004715,0.70197983){\color[rgb]{0,0,0}\makebox(0,0)[t]{\lineheight{1.25}\smash{\begin{tabular}[t]{c}\textbf{\stateafl}\end{tabular}}}}%
    \put(0,0){\includegraphics[width=\unitlength,page=4]{group_1.pdf}}%
    \put(0.67552251,0.70265821){\color[rgb]{0,0,0}\makebox(0,0)[t]{\lineheight{1.25}\smash{\begin{tabular}[t]{c}\textbf{\sgfuzz}\end{tabular}}}}%
    \put(0,0){\includegraphics[width=\unitlength,page=5]{group_1.pdf}}%
    \put(0.09842848,0.49135943){\color[rgb]{0,0,0}\makebox(0,0)[lt]{\lineheight{1.25}\smash{\begin{tabular}[t]{l}00:00\end{tabular}}}}%
    \put(0.1302572,0.49135943){\color[rgb]{0,0,0}\transparent{0}\makebox(0,0)[t]{\lineheight{1.25}\smash{\begin{tabular}[t]{c}03:00\end{tabular}}}}%
    \put(0.16208592,0.49135943){\color[rgb]{0,0,0}\makebox(0,0)[t]{\lineheight{1.25}\smash{\begin{tabular}[t]{c}06:00\end{tabular}}}}%
    \put(0.19391465,0.49135943){\color[rgb]{0,0,0}\transparent{0}\makebox(0,0)[t]{\lineheight{1.25}\smash{\begin{tabular}[t]{c}09:00\end{tabular}}}}%
    \put(0.22574335,0.49135943){\color[rgb]{0,0,0}\makebox(0,0)[t]{\lineheight{1.25}\smash{\begin{tabular}[t]{c}12:00\end{tabular}}}}%
    \put(0.25757209,0.49135943){\color[rgb]{0,0,0}\transparent{0}\makebox(0,0)[t]{\lineheight{1.25}\smash{\begin{tabular}[t]{c}15:00\end{tabular}}}}%
    \put(0.28940081,0.49135943){\color[rgb]{0,0,0}\makebox(0,0)[t]{\lineheight{1.25}\smash{\begin{tabular}[t]{c}18:00\end{tabular}}}}%
    \put(0.32122953,0.49135943){\color[rgb]{0,0,0}\transparent{0}\makebox(0,0)[t]{\lineheight{1.25}\smash{\begin{tabular}[t]{c}21:00\end{tabular}}}}%
    \put(0,0){\includegraphics[width=\unitlength,page=6]{group_1.pdf}}%
    \put(0.09249125,0.50153755){\color[rgb]{0,0,0}\makebox(0,0)[rt]{\lineheight{1.25}\smash{\begin{tabular}[t]{r}0\end{tabular}}}}%
    \put(0.09249125,0.53687824){\color[rgb]{0,0,0}\makebox(0,0)[rt]{\lineheight{1.25}\smash{\begin{tabular}[t]{r}500\end{tabular}}}}%
    \put(0.09249125,0.57221891){\color[rgb]{0,0,0}\makebox(0,0)[rt]{\lineheight{1.25}\smash{\begin{tabular}[t]{r}1,000\end{tabular}}}}%
    \put(0.09249125,0.6075596){\color[rgb]{0,0,0}\makebox(0,0)[rt]{\lineheight{1.25}\smash{\begin{tabular}[t]{r}1,500\end{tabular}}}}%
    \put(0.09249125,0.64290029){\color[rgb]{0,0,0}\makebox(0,0)[rt]{\lineheight{1.25}\smash{\begin{tabular}[t]{r}2,000\end{tabular}}}}%
    \put(0,0){\includegraphics[width=\unitlength,page=7]{group_1.pdf}}%
    \put(0.22523086,0.67711007){\color[rgb]{0,0,0}\makebox(0,0)[t]{\lineheight{1.25}\smash{\begin{tabular}[t]{c}\textbf{\dropbearServer}\end{tabular}}}}%
    \put(0.39402506,0.11143884){\color[rgb]{0,0,0}\makebox(0,0)[rt]{\lineheight{1.25}\smash{\begin{tabular}[t]{r}1,000\end{tabular}}}}%
    \put(0.39402506,0.14913557){\color[rgb]{0,0,0}\makebox(0,0)[rt]{\lineheight{1.25}\smash{\begin{tabular}[t]{r}2,000\end{tabular}}}}%
    \put(0.39402506,0.1868323){\color[rgb]{0,0,0}\makebox(0,0)[rt]{\lineheight{1.25}\smash{\begin{tabular}[t]{r}3,000\end{tabular}}}}%
    \put(0.39402506,0.22452903){\color[rgb]{0,0,0}\makebox(0,0)[rt]{\lineheight{1.25}\smash{\begin{tabular}[t]{r}4,000\end{tabular}}}}%
    \put(0,0){\includegraphics[width=\unitlength,page=8]{group_1.pdf}}%
    \put(0.39672697,0.06356399){\color[rgb]{0,0,0}\makebox(0,0)[lt]{\lineheight{1.25}\smash{\begin{tabular}[t]{l}00:00\end{tabular}}}}%
    \put(0.42855569,0.06356399){\color[rgb]{0,0,0}\transparent{0}\makebox(0,0)[t]{\lineheight{1.25}\smash{\begin{tabular}[t]{c}03:00\end{tabular}}}}%
    \put(0.46038441,0.06356399){\color[rgb]{0,0,0}\makebox(0,0)[t]{\lineheight{1.25}\smash{\begin{tabular}[t]{c}06:00\end{tabular}}}}%
    \put(0.49221313,0.06356399){\color[rgb]{0,0,0}\transparent{0}\makebox(0,0)[t]{\lineheight{1.25}\smash{\begin{tabular}[t]{c}09:00\end{tabular}}}}%
    \put(0.52404185,0.06356399){\color[rgb]{0,0,0}\makebox(0,0)[t]{\lineheight{1.25}\smash{\begin{tabular}[t]{c}12:00\end{tabular}}}}%
    \put(0.55587057,0.06356399){\color[rgb]{0,0,0}\transparent{0}\makebox(0,0)[t]{\lineheight{1.25}\smash{\begin{tabular}[t]{c}15:00\end{tabular}}}}%
    \put(0.58769929,0.06356399){\color[rgb]{0,0,0}\makebox(0,0)[t]{\lineheight{1.25}\smash{\begin{tabular}[t]{c}18:00\end{tabular}}}}%
    \put(0.61952801,0.06356399){\color[rgb]{0,0,0}\transparent{0}\makebox(0,0)[t]{\lineheight{1.25}\smash{\begin{tabular}[t]{c}21:00\end{tabular}}}}%
    \put(0,0){\includegraphics[width=\unitlength,page=9]{group_1.pdf}}%
    \put(0.39078973,0.07374211){\color[rgb]{0,0,0}\makebox(0,0)[rt]{\lineheight{1.25}\smash{\begin{tabular}[t]{r}0\end{tabular}}}}%
    \put(0,0){\includegraphics[width=\unitlength,page=10]{group_1.pdf}}%
    \put(0.52352934,0.24867933){\color[rgb]{0,0,0}\makebox(0,0)[t]{\lineheight{1.25}\smash{\begin{tabular}[t]{c}\textbf{\dcmtkServer}\end{tabular}}}}%
    \put(0,0){\includegraphics[width=\unitlength,page=11]{group_1.pdf}}%
    \put(0.09863556,0.27661354){\color[rgb]{0,0,0}\makebox(0,0)[lt]{\lineheight{1.25}\smash{\begin{tabular}[t]{l}00:00\end{tabular}}}}%
    \put(0.13046428,0.27661354){\color[rgb]{0,0,0}\transparent{0}\makebox(0,0)[t]{\lineheight{1.25}\smash{\begin{tabular}[t]{c}03:00\end{tabular}}}}%
    \put(0.162293,0.27661354){\color[rgb]{0,0,0}\makebox(0,0)[t]{\lineheight{1.25}\smash{\begin{tabular}[t]{c}06:00\end{tabular}}}}%
    \put(0.19412172,0.27661354){\color[rgb]{0,0,0}\transparent{0}\makebox(0,0)[t]{\lineheight{1.25}\smash{\begin{tabular}[t]{c}09:00\end{tabular}}}}%
    \put(0.22595044,0.27661354){\color[rgb]{0,0,0}\makebox(0,0)[t]{\lineheight{1.25}\smash{\begin{tabular}[t]{c}12:00\end{tabular}}}}%
    \put(0.25777916,0.27661354){\color[rgb]{0,0,0}\transparent{0}\makebox(0,0)[t]{\lineheight{1.25}\smash{\begin{tabular}[t]{c}15:00\end{tabular}}}}%
    \put(0.28960788,0.27661354){\color[rgb]{0,0,0}\makebox(0,0)[t]{\lineheight{1.25}\smash{\begin{tabular}[t]{c}18:00\end{tabular}}}}%
    \put(0.32143655,0.27661354){\color[rgb]{0,0,0}\transparent{0}\makebox(0,0)[t]{\lineheight{1.25}\smash{\begin{tabular}[t]{c}21:00\end{tabular}}}}%
    \put(0,0){\includegraphics[width=\unitlength,page=12]{group_1.pdf}}%
    \put(0.09269832,0.28679165){\color[rgb]{0,0,0}\makebox(0,0)[rt]{\lineheight{1.25}\smash{\begin{tabular}[t]{r}0\end{tabular}}}}%
    \put(0.09269832,0.3150642){\color[rgb]{0,0,0}\makebox(0,0)[rt]{\lineheight{1.25}\smash{\begin{tabular}[t]{r}1,000\end{tabular}}}}%
    \put(0.09269832,0.34333675){\color[rgb]{0,0,0}\makebox(0,0)[rt]{\lineheight{1.25}\smash{\begin{tabular}[t]{r}2,000\end{tabular}}}}%
    \put(0.09269832,0.3716093){\color[rgb]{0,0,0}\makebox(0,0)[rt]{\lineheight{1.25}\smash{\begin{tabular}[t]{r}3,000\end{tabular}}}}%
    \put(0.09269832,0.39988184){\color[rgb]{0,0,0}\makebox(0,0)[rt]{\lineheight{1.25}\smash{\begin{tabular}[t]{r}4,000\end{tabular}}}}%
    \put(0.09269832,0.4281544){\color[rgb]{0,0,0}\makebox(0,0)[rt]{\lineheight{1.25}\smash{\begin{tabular}[t]{r}5,000\end{tabular}}}}%
    \put(0.09269832,0.45642694){\color[rgb]{0,0,0}\makebox(0,0)[rt]{\lineheight{1.25}\smash{\begin{tabular}[t]{r}6,000\end{tabular}}}}%
    \put(0,0){\includegraphics[width=\unitlength,page=13]{group_1.pdf}}%
    \put(0.22543793,0.46240724){\color[rgb]{0,0,0}\makebox(0,0)[t]{\lineheight{1.25}\smash{\begin{tabular}[t]{c}\textbf{\gnutlsServer}\end{tabular}}}}%
    \put(0,0){\includegraphics[width=\unitlength,page=14]{group_1.pdf}}%
    \put(0.39901109,0.27938716){\color[rgb]{0,0,0}\makebox(0,0)[lt]{\lineheight{1.25}\smash{\begin{tabular}[t]{l}00:00\end{tabular}}}}%
    \put(0.43083982,0.27938716){\color[rgb]{0,0,0}\transparent{0}\makebox(0,0)[t]{\lineheight{1.25}\smash{\begin{tabular}[t]{c}03:00\end{tabular}}}}%
    \put(0.46266854,0.27938716){\color[rgb]{0,0,0}\makebox(0,0)[t]{\lineheight{1.25}\smash{\begin{tabular}[t]{c}06:00\end{tabular}}}}%
    \put(0.49449726,0.27938716){\color[rgb]{0,0,0}\transparent{0}\makebox(0,0)[t]{\lineheight{1.25}\smash{\begin{tabular}[t]{c}09:00\end{tabular}}}}%
    \put(0.52632597,0.27938716){\color[rgb]{0,0,0}\makebox(0,0)[t]{\lineheight{1.25}\smash{\begin{tabular}[t]{c}12:00\end{tabular}}}}%
    \put(0.5581547,0.27938716){\color[rgb]{0,0,0}\transparent{0}\makebox(0,0)[t]{\lineheight{1.25}\smash{\begin{tabular}[t]{c}15:00\end{tabular}}}}%
    \put(0.58998342,0.27938716){\color[rgb]{0,0,0}\makebox(0,0)[t]{\lineheight{1.25}\smash{\begin{tabular}[t]{c}18:00\end{tabular}}}}%
    \put(0.62181214,0.27938716){\color[rgb]{0,0,0}\transparent{0}\makebox(0,0)[t]{\lineheight{1.25}\smash{\begin{tabular}[t]{c}21:00\end{tabular}}}}%
    \put(0,0){\includegraphics[width=\unitlength,page=15]{group_1.pdf}}%
    \put(0.39307386,0.28956527){\color[rgb]{0,0,0}\makebox(0,0)[rt]{\lineheight{1.25}\smash{\begin{tabular}[t]{r}0\end{tabular}}}}%
    \put(0.39307386,0.33197409){\color[rgb]{0,0,0}\makebox(0,0)[rt]{\lineheight{1.25}\smash{\begin{tabular}[t]{r}2,000\end{tabular}}}}%
    \put(0.39307386,0.37438291){\color[rgb]{0,0,0}\makebox(0,0)[rt]{\lineheight{1.25}\smash{\begin{tabular}[t]{r}4,000\end{tabular}}}}%
    \put(0.39307386,0.41679174){\color[rgb]{0,0,0}\makebox(0,0)[rt]{\lineheight{1.25}\smash{\begin{tabular}[t]{r}6,000\end{tabular}}}}%
    \put(0.39307386,0.45920056){\color[rgb]{0,0,0}\makebox(0,0)[rt]{\lineheight{1.25}\smash{\begin{tabular}[t]{r}8,000\end{tabular}}}}%
    \put(0,0){\includegraphics[width=\unitlength,page=16]{group_1.pdf}}%
    \put(0.52581347,0.46690442){\color[rgb]{0,0,0}\makebox(0,0)[t]{\lineheight{1.25}\smash{\begin{tabular}[t]{c}\textbf{\libresslServer}\end{tabular}}}}%
    \put(0,0){\includegraphics[width=\unitlength,page=17]{group_1.pdf}}%
    \put(0.69984108,0.06356399){\color[rgb]{0,0,0}\makebox(0,0)[lt]{\lineheight{1.25}\smash{\begin{tabular}[t]{l}00:00\end{tabular}}}}%
    \put(0.7316698,0.06356399){\color[rgb]{0,0,0}\transparent{0}\makebox(0,0)[t]{\lineheight{1.25}\smash{\begin{tabular}[t]{c}03:00\end{tabular}}}}%
    \put(0.76349852,0.06356399){\color[rgb]{0,0,0}\makebox(0,0)[t]{\lineheight{1.25}\smash{\begin{tabular}[t]{c}06:00\end{tabular}}}}%
    \put(0.79532724,0.06356399){\color[rgb]{0,0,0}\transparent{0}\makebox(0,0)[t]{\lineheight{1.25}\smash{\begin{tabular}[t]{c}09:00\end{tabular}}}}%
    \put(0.82715596,0.06356399){\color[rgb]{0,0,0}\makebox(0,0)[t]{\lineheight{1.25}\smash{\begin{tabular}[t]{c}12:00\end{tabular}}}}%
    \put(0.85898468,0.06356399){\color[rgb]{0,0,0}\transparent{0}\makebox(0,0)[t]{\lineheight{1.25}\smash{\begin{tabular}[t]{c}15:00\end{tabular}}}}%
    \put(0.8908134,0.06356399){\color[rgb]{0,0,0}\makebox(0,0)[t]{\lineheight{1.25}\smash{\begin{tabular}[t]{c}18:00\end{tabular}}}}%
    \put(0.92264212,0.06356399){\color[rgb]{0,0,0}\transparent{0}\makebox(0,0)[t]{\lineheight{1.25}\smash{\begin{tabular}[t]{c}21:00\end{tabular}}}}%
    \put(0,0){\includegraphics[width=\unitlength,page=18]{group_1.pdf}}%
    \put(0.69390384,0.07374211){\color[rgb]{0,0,0}\makebox(0,0)[rt]{\lineheight{1.25}\smash{\begin{tabular}[t]{r}0\end{tabular}}}}%
    \put(0.69390384,0.12086304){\color[rgb]{0,0,0}\makebox(0,0)[rt]{\lineheight{1.25}\smash{\begin{tabular}[t]{r}500\end{tabular}}}}%
    \put(0.69390384,0.16798394){\color[rgb]{0,0,0}\makebox(0,0)[rt]{\lineheight{1.25}\smash{\begin{tabular}[t]{r}1,000\end{tabular}}}}%
    \put(0.69390384,0.21510484){\color[rgb]{0,0,0}\makebox(0,0)[rt]{\lineheight{1.25}\smash{\begin{tabular}[t]{r}1,500\end{tabular}}}}%
    \put(0,0){\includegraphics[width=\unitlength,page=19]{group_1.pdf}}%
    \put(0.82664345,0.25108126){\color[rgb]{0,0,0}\makebox(0,0)[t]{\lineheight{1.25}\smash{\begin{tabular}[t]{c}\textbf{\livefivefivefiveServer}\end{tabular}}}}%
    \put(0,0){\includegraphics[width=\unitlength,page=20]{group_1.pdf}}%
    \put(0.09842848,0.06356399){\color[rgb]{0,0,0}\makebox(0,0)[lt]{\lineheight{1.25}\smash{\begin{tabular}[t]{l}00:00\end{tabular}}}}%
    \put(0.1302572,0.06356399){\color[rgb]{0,0,0}\transparent{0}\makebox(0,0)[t]{\lineheight{1.25}\smash{\begin{tabular}[t]{c}03:00\end{tabular}}}}%
    \put(0.16208592,0.06356399){\color[rgb]{0,0,0}\makebox(0,0)[t]{\lineheight{1.25}\smash{\begin{tabular}[t]{c}06:00\end{tabular}}}}%
    \put(0.19391464,0.06356399){\color[rgb]{0,0,0}\transparent{0}\makebox(0,0)[t]{\lineheight{1.25}\smash{\begin{tabular}[t]{c}09:00\end{tabular}}}}%
    \put(0.22574337,0.06356399){\color[rgb]{0,0,0}\makebox(0,0)[t]{\lineheight{1.25}\smash{\begin{tabular}[t]{c}12:00\end{tabular}}}}%
    \put(0.25757203,0.06356399){\color[rgb]{0,0,0}\transparent{0}\makebox(0,0)[t]{\lineheight{1.25}\smash{\begin{tabular}[t]{c}15:00\end{tabular}}}}%
    \put(0.28940081,0.06356399){\color[rgb]{0,0,0}\makebox(0,0)[t]{\lineheight{1.25}\smash{\begin{tabular}[t]{c}18:00\end{tabular}}}}%
    \put(0.32122948,0.06356399){\color[rgb]{0,0,0}\transparent{0}\makebox(0,0)[t]{\lineheight{1.25}\smash{\begin{tabular}[t]{c}21:00\end{tabular}}}}%
    \put(0,0){\includegraphics[width=\unitlength,page=21]{group_1.pdf}}%
    \put(0.09249125,0.07374211){\color[rgb]{0,0,0}\makebox(0,0)[rt]{\lineheight{1.25}\smash{\begin{tabular}[t]{r}0\end{tabular}}}}%
    \put(0.09249125,0.1122956){\color[rgb]{0,0,0}\makebox(0,0)[rt]{\lineheight{1.25}\smash{\begin{tabular}[t]{r}500\end{tabular}}}}%
    \put(0.09249125,0.15084906){\color[rgb]{0,0,0}\makebox(0,0)[rt]{\lineheight{1.25}\smash{\begin{tabular}[t]{r}1,000\end{tabular}}}}%
    \put(0.09249125,0.18940254){\color[rgb]{0,0,0}\makebox(0,0)[rt]{\lineheight{1.25}\smash{\begin{tabular}[t]{r}1,500\end{tabular}}}}%
    \put(0.09249125,0.227956){\color[rgb]{0,0,0}\makebox(0,0)[rt]{\lineheight{1.25}\smash{\begin{tabular}[t]{r}2,000\end{tabular}}}}%
    \put(0,0){\includegraphics[width=\unitlength,page=22]{group_1.pdf}}%
    \put(0.23655891,0.25171656){\color[rgb]{0,0,0}\makebox(0,0)[t]{\lineheight{1.25}\smash{\begin{tabular}[t]{c}\textbf{\mosquittoServer}\end{tabular}}}}%
    \put(0,0){\includegraphics[width=\unitlength,page=23]{group_1.pdf}}%
    \put(0.39841692,0.49135943){\color[rgb]{0,0,0}\makebox(0,0)[lt]{\lineheight{1.25}\smash{\begin{tabular}[t]{l}00:00\end{tabular}}}}%
    \put(0.43024564,0.49135943){\color[rgb]{0,0,0}\transparent{0}\makebox(0,0)[t]{\lineheight{1.25}\smash{\begin{tabular}[t]{c}03:00\end{tabular}}}}%
    \put(0.46207436,0.49135943){\color[rgb]{0,0,0}\makebox(0,0)[t]{\lineheight{1.25}\smash{\begin{tabular}[t]{c}06:00\end{tabular}}}}%
    \put(0.49390308,0.49135943){\color[rgb]{0,0,0}\transparent{0}\makebox(0,0)[t]{\lineheight{1.25}\smash{\begin{tabular}[t]{c}09:00\end{tabular}}}}%
    \put(0.5257318,0.49135943){\color[rgb]{0,0,0}\makebox(0,0)[t]{\lineheight{1.25}\smash{\begin{tabular}[t]{c}12:00\end{tabular}}}}%
    \put(0.55756052,0.49135943){\color[rgb]{0,0,0}\transparent{0}\makebox(0,0)[t]{\lineheight{1.25}\smash{\begin{tabular}[t]{c}15:00\end{tabular}}}}%
    \put(0.58938924,0.49135943){\color[rgb]{0,0,0}\makebox(0,0)[t]{\lineheight{1.25}\smash{\begin{tabular}[t]{c}18:00\end{tabular}}}}%
    \put(0.62121796,0.49135943){\color[rgb]{0,0,0}\transparent{0}\makebox(0,0)[t]{\lineheight{1.25}\smash{\begin{tabular}[t]{c}21:00\end{tabular}}}}%
    \put(0,0){\includegraphics[width=\unitlength,page=24]{group_1.pdf}}%
    \put(0.39247969,0.50153755){\color[rgb]{0,0,0}\makebox(0,0)[rt]{\lineheight{1.25}\smash{\begin{tabular}[t]{r}0\end{tabular}}}}%
    \put(0.39247969,0.53238034){\color[rgb]{0,0,0}\makebox(0,0)[rt]{\lineheight{1.25}\smash{\begin{tabular}[t]{r}1,000\end{tabular}}}}%
    \put(0.39247969,0.56322313){\color[rgb]{0,0,0}\makebox(0,0)[rt]{\lineheight{1.25}\smash{\begin{tabular}[t]{r}2,000\end{tabular}}}}%
    \put(0.39247969,0.59406586){\color[rgb]{0,0,0}\makebox(0,0)[rt]{\lineheight{1.25}\smash{\begin{tabular}[t]{r}3,000\end{tabular}}}}%
    \put(0.39247969,0.62490865){\color[rgb]{0,0,0}\makebox(0,0)[rt]{\lineheight{1.25}\smash{\begin{tabular}[t]{r}4,000\end{tabular}}}}%
    \put(0.39247969,0.65575144){\color[rgb]{0,0,0}\makebox(0,0)[rt]{\lineheight{1.25}\smash{\begin{tabular}[t]{r}5,000\end{tabular}}}}%
    \put(0,0){\includegraphics[width=\unitlength,page=25]{group_1.pdf}}%
    \put(0.5252193,0.67880642){\color[rgb]{0,0,0}\makebox(0,0)[t]{\lineheight{1.25}\smash{\begin{tabular}[t]{c}\textbf{\nginxServer}\end{tabular}}}}%
    \put(0,0){\includegraphics[width=\unitlength,page=26]{group_1.pdf}}%
    \put(0.70029558,0.27873398){\color[rgb]{0,0,0}\makebox(0,0)[lt]{\lineheight{1.25}\smash{\begin{tabular}[t]{l}00:00\end{tabular}}}}%
    \put(0.7321243,0.27873398){\color[rgb]{0,0,0}\transparent{0}\makebox(0,0)[t]{\lineheight{1.25}\smash{\begin{tabular}[t]{c}03:00\end{tabular}}}}%
    \put(0.76395302,0.27873398){\color[rgb]{0,0,0}\makebox(0,0)[t]{\lineheight{1.25}\smash{\begin{tabular}[t]{c}06:00\end{tabular}}}}%
    \put(0.79578174,0.27873398){\color[rgb]{0,0,0}\transparent{0}\makebox(0,0)[t]{\lineheight{1.25}\smash{\begin{tabular}[t]{c}09:00\end{tabular}}}}%
    \put(0.82761046,0.27873398){\color[rgb]{0,0,0}\makebox(0,0)[t]{\lineheight{1.25}\smash{\begin{tabular}[t]{c}12:00\end{tabular}}}}%
    \put(0.85943918,0.27873398){\color[rgb]{0,0,0}\transparent{0}\makebox(0,0)[t]{\lineheight{1.25}\smash{\begin{tabular}[t]{c}15:00\end{tabular}}}}%
    \put(0.8912679,0.27873398){\color[rgb]{0,0,0}\makebox(0,0)[t]{\lineheight{1.25}\smash{\begin{tabular}[t]{c}18:00\end{tabular}}}}%
    \put(0.92309662,0.27873398){\color[rgb]{0,0,0}\transparent{0}\makebox(0,0)[t]{\lineheight{1.25}\smash{\begin{tabular}[t]{c}21:00\end{tabular}}}}%
    \put(0,0){\includegraphics[width=\unitlength,page=27]{group_1.pdf}}%
    \put(0.69435835,0.28891209){\color[rgb]{0,0,0}\makebox(0,0)[rt]{\lineheight{1.25}\smash{\begin{tabular}[t]{r}0\end{tabular}}}}%
    \put(0.69435835,0.34192312){\color[rgb]{0,0,0}\makebox(0,0)[rt]{\lineheight{1.25}\smash{\begin{tabular}[t]{r}5,000\end{tabular}}}}%
    \put(0.69435835,0.39493415){\color[rgb]{0,0,0}\makebox(0,0)[rt]{\lineheight{1.25}\smash{\begin{tabular}[t]{r}10,000\end{tabular}}}}%
    \put(0.69435835,0.44794517){\color[rgb]{0,0,0}\makebox(0,0)[rt]{\lineheight{1.25}\smash{\begin{tabular}[t]{r}15,000\end{tabular}}}}%
    \put(0,0){\includegraphics[width=\unitlength,page=28]{group_1.pdf}}%
    \put(0.82709796,0.4661379){\color[rgb]{0,0,0}\makebox(0,0)[t]{\lineheight{1.25}\smash{\begin{tabular}[t]{c}\textbf{\opensslServer}\end{tabular}}}}%
    \put(0,0){\includegraphics[width=\unitlength,page=29]{group_1.pdf}}%
    \put(0.70074116,0.49238694){\color[rgb]{0,0,0}\makebox(0,0)[lt]{\lineheight{1.25}\smash{\begin{tabular}[t]{l}00:00\end{tabular}}}}%
    \put(0.73256988,0.49238694){\color[rgb]{0,0,0}\transparent{0}\makebox(0,0)[t]{\lineheight{1.25}\smash{\begin{tabular}[t]{c}03:00\end{tabular}}}}%
    \put(0.7643986,0.49238694){\color[rgb]{0,0,0}\makebox(0,0)[t]{\lineheight{1.25}\smash{\begin{tabular}[t]{c}06:00\end{tabular}}}}%
    \put(0.79622731,0.49238694){\color[rgb]{0,0,0}\transparent{0}\makebox(0,0)[t]{\lineheight{1.25}\smash{\begin{tabular}[t]{c}09:00\end{tabular}}}}%
    \put(0.82805603,0.49238694){\color[rgb]{0,0,0}\makebox(0,0)[t]{\lineheight{1.25}\smash{\begin{tabular}[t]{c}12:00\end{tabular}}}}%
    \put(0.85988475,0.49238694){\color[rgb]{0,0,0}\transparent{0}\makebox(0,0)[t]{\lineheight{1.25}\smash{\begin{tabular}[t]{c}15:00\end{tabular}}}}%
    \put(0.89171347,0.49238694){\color[rgb]{0,0,0}\makebox(0,0)[t]{\lineheight{1.25}\smash{\begin{tabular}[t]{c}18:00\end{tabular}}}}%
    \put(0.92354219,0.49238694){\color[rgb]{0,0,0}\transparent{0}\makebox(0,0)[t]{\lineheight{1.25}\smash{\begin{tabular}[t]{c}21:00\end{tabular}}}}%
    \put(0,0){\includegraphics[width=\unitlength,page=30]{group_1.pdf}}%
    \put(0.69480392,0.50256506){\color[rgb]{0,0,0}\makebox(0,0)[rt]{\lineheight{1.25}\smash{\begin{tabular}[t]{r}0\end{tabular}}}}%
    \put(0.69480392,0.53649211){\color[rgb]{0,0,0}\makebox(0,0)[rt]{\lineheight{1.25}\smash{\begin{tabular}[t]{r}2,000\end{tabular}}}}%
    \put(0.69480392,0.57041917){\color[rgb]{0,0,0}\makebox(0,0)[rt]{\lineheight{1.25}\smash{\begin{tabular}[t]{r}4,000\end{tabular}}}}%
    \put(0.69480392,0.60434623){\color[rgb]{0,0,0}\makebox(0,0)[rt]{\lineheight{1.25}\smash{\begin{tabular}[t]{r}6,000\end{tabular}}}}%
    \put(0.69480392,0.63827328){\color[rgb]{0,0,0}\makebox(0,0)[rt]{\lineheight{1.25}\smash{\begin{tabular}[t]{r}8,000\end{tabular}}}}%
    \put(0.69480392,0.67220034){\color[rgb]{0,0,0}\makebox(0,0)[rt]{\lineheight{1.25}\smash{\begin{tabular}[t]{r}10,000\end{tabular}}}}%
    \put(0,0){\includegraphics[width=\unitlength,page=31]{group_1.pdf}}%
    \put(0.82754353,0.67915555){\color[rgb]{0,0,0}\makebox(0,0)[t]{\lineheight{1.25}\smash{\begin{tabular}[t]{c}\textbf{\sambaServer}\end{tabular}}}}%
  \end{picture}%
\endgroup%

%% file: eval-plots/mosquitto_tls_with_diff.pdf_tex
\begingroup%
  \makeatletter%
  \providecommand\color[2][]{%
    \errmessage{(Inkscape) Color is used for the text in Inkscape, but the package 'color.sty' is not loaded}%
    \renewcommand\color[2][]{}%
  }%
  \providecommand\transparent[1]{%
    \errmessage{(Inkscape) Transparency is used (non-zero) for the text in Inkscape, but the package 'transparent.sty' is not loaded}%
    \renewcommand\transparent[1]{}%
  }%
  \providecommand\rotatebox[2]{#2}%
  \newcommand*\fsize{\dimexpr\f@size pt\relax}%
  \newcommand*\lineheight[1]{\fontsize{\fsize}{#1\fsize}\selectfont}%
  \ifx\svgwidth\undefined%
    \setlength{\unitlength}{270bp}%
    \ifx\svgscale\undefined%
      \relax%
    \else%
      \setlength{\unitlength}{\unitlength * \real{\svgscale}}%
    \fi%
  \else%
    \setlength{\unitlength}{\svgwidth}%
  \fi%
  \global\let\svgwidth\undefined%
  \global\let\svgscale\undefined%
  \makeatother%
  \begin{picture}(1,0.85)%
    \lineheight{1}%
    \setlength\tabcolsep{0pt}%
    \put(0,0){\includegraphics[width=\unitlength,page=1]{mosquitto_tls_with_diff.pdf}}%
    \put(0.14861111,0.05972222){\color[rgb]{0,0,0}\makebox(0,0)[lt]{\lineheight{1.25}\smash{\begin{tabular}[t]{l}00:00\end{tabular}}}}%
    \put(0.25285017,0.05972222){\color[rgb]{0,0,0}\transparent{0}\makebox(0,0)[t]{\lineheight{1.25}\smash{\begin{tabular}[t]{c}03:00\end{tabular}}}}%
    \put(0.35708923,0.05972222){\color[rgb]{0,0,0}\makebox(0,0)[t]{\lineheight{1.25}\smash{\begin{tabular}[t]{c}06:00\end{tabular}}}}%
    \put(0.46132829,0.05972222){\color[rgb]{0,0,0}\transparent{0}\makebox(0,0)[t]{\lineheight{1.25}\smash{\begin{tabular}[t]{c}09:00\end{tabular}}}}%
    \put(0.56556731,0.05972222){\color[rgb]{0,0,0}\makebox(0,0)[t]{\lineheight{1.25}\smash{\begin{tabular}[t]{c}12:00\end{tabular}}}}%
    \put(0.66980637,0.05972222){\color[rgb]{0,0,0}\transparent{0}\makebox(0,0)[t]{\lineheight{1.25}\smash{\begin{tabular}[t]{c}15:00\end{tabular}}}}%
    \put(0.77404548,0.05972222){\color[rgb]{0,0,0}\makebox(0,0)[t]{\lineheight{1.25}\smash{\begin{tabular}[t]{c}18:00\end{tabular}}}}%
    \put(0.87828454,0.05972222){\color[rgb]{0,0,0}\transparent{0}\makebox(0,0)[t]{\lineheight{1.25}\smash{\begin{tabular}[t]{c}21:00\end{tabular}}}}%
    \put(0,0){\includegraphics[width=\unitlength,page=2]{mosquitto_tls_with_diff.pdf}}%
    \put(0.56527778,0.01805556){\color[rgb]{0,0,0}\makebox(0,0)[t]{\lineheight{1.25}\smash{\begin{tabular}[t]{c}\textbf{Time [hh:mm]}\end{tabular}}}}%
    \put(0,0){\includegraphics[width=\unitlength,page=3]{mosquitto_tls_with_diff.pdf}}%
    \put(0.12916667,0.09305556){\color[rgb]{0,0,0}\makebox(0,0)[rt]{\lineheight{1.25}\smash{\begin{tabular}[t]{r}0\end{tabular}}}}%
    \put(0.12916667,0.21995265){\color[rgb]{0,0,0}\makebox(0,0)[rt]{\lineheight{1.25}\smash{\begin{tabular}[t]{r}500\end{tabular}}}}%
    \put(0.12916667,0.34684978){\color[rgb]{0,0,0}\makebox(0,0)[rt]{\lineheight{1.25}\smash{\begin{tabular}[t]{r}1,000\end{tabular}}}}%
    \put(0.12916667,0.47374687){\color[rgb]{0,0,0}\makebox(0,0)[rt]{\lineheight{1.25}\smash{\begin{tabular}[t]{r}1,500\end{tabular}}}}%
    \put(0.12916667,0.60064401){\color[rgb]{0,0,0}\makebox(0,0)[rt]{\lineheight{1.25}\smash{\begin{tabular}[t]{r}2,000\end{tabular}}}}%
    \put(0,0){\includegraphics[width=\unitlength,page=4]{mosquitto_tls_with_diff.pdf}}%
    \put(0.04298774,0.37916667){\color[rgb]{0,0,0}\rotatebox{90}{\makebox(0,0)[t]{\lineheight{1.25}\smash{\begin{tabular}[t]{c}\textbf{\#Covered Branches}\end{tabular}}}}}%
    \put(0,0){\includegraphics[width=\unitlength,page=5]{mosquitto_tls_with_diff.pdf}}%
    \put(0.72592238,0.67581245){\color[rgb]{0,0,0}\makebox(0,0)[lt]{\lineheight{1.25}\smash{\begin{tabular}[t]{l}enabled\end{tabular}}}}%
    \put(0,0){\includegraphics[width=\unitlength,page=6]{mosquitto_tls_with_diff.pdf}}%
    \put(0.86802953,0.67581245){\color[rgb]{0,0,0}\makebox(0,0)[lt]{\lineheight{1.25}\smash{\begin{tabular}[t]{l}disabled\end{tabular}}}}%
    \put(0.68486727,0.70548986){\color[rgb]{0,0,0}\makebox(0,0)[lt]{\lineheight{1.25}\smash{\begin{tabular}[t]{l}\textbf{TLS}\end{tabular}}}}%
    \put(0.15231112,0.70569127){\color[rgb]{0,0,0}\makebox(0,0)[lt]{\lineheight{1.25}\smash{\begin{tabular}[t]{l}\textbf{Fuzzer}\end{tabular}}}}%
    \put(0.53603817,0.73582868){\color[rgb]{0,0,0}\makebox(0,0)[t]{\lineheight{1.25}\smash{\begin{tabular}[t]{c}\textbf{\mosquittoServer}\end{tabular}}}}%
    \put(0.18647799,0.67662656){\color[rgb]{0,0,0}\makebox(0,0)[lt]{\lineheight{1.25}\smash{\begin{tabular}[t]{l}\textbf{\toolname}\end{tabular}}}}%
    \put(0,0){\includegraphics[width=\unitlength,page=7]{mosquitto_tls_with_diff.pdf}}%
    \put(0.37325516,0.67790082){\color[rgb]{0,0,0}\makebox(0,0)[lt]{\lineheight{1.25}\smash{\begin{tabular}[t]{l}\textbf{\sgfuzz}\end{tabular}}}}%
    \put(0,0){\includegraphics[width=\unitlength,page=8]{mosquitto_tls_with_diff.pdf}}%
  \end{picture}%
\endgroup%

%% file: eval-plots/live555_auth_with_diff.pdf_tex
\begingroup%
  \makeatletter%
  \providecommand\color[2][]{%
    \errmessage{(Inkscape) Color is used for the text in Inkscape, but the package 'color.sty' is not loaded}%
    \renewcommand\color[2][]{}%
  }%
  \providecommand\transparent[1]{%
    \errmessage{(Inkscape) Transparency is used (non-zero) for the text in Inkscape, but the package 'transparent.sty' is not loaded}%
    \renewcommand\transparent[1]{}%
  }%
  \providecommand\rotatebox[2]{#2}%
  \newcommand*\fsize{\dimexpr\f@size pt\relax}%
  \newcommand*\lineheight[1]{\fontsize{\fsize}{#1\fsize}\selectfont}%
  \ifx\svgwidth\undefined%
    \setlength{\unitlength}{270bp}%
    \ifx\svgscale\undefined%
      \relax%
    \else%
      \setlength{\unitlength}{\unitlength * \real{\svgscale}}%
    \fi%
  \else%
    \setlength{\unitlength}{\svgwidth}%
  \fi%
  \global\let\svgwidth\undefined%
  \global\let\svgscale\undefined%
  \makeatother%
  \begin{picture}(1,0.85)%
    \lineheight{1}%
    \setlength\tabcolsep{0pt}%
    \put(0,0){\includegraphics[width=\unitlength,page=1]{live555_auth_with_diff.pdf}}%
    \put(0.14861111,0.05972222){\color[rgb]{0,0,0}\makebox(0,0)[lt]{\lineheight{1.25}\smash{\begin{tabular}[t]{l}00:00\end{tabular}}}}%
    \put(0.25285017,0.05972222){\color[rgb]{0,0,0}\transparent{0}\makebox(0,0)[t]{\lineheight{1.25}\smash{\begin{tabular}[t]{c}03:00\end{tabular}}}}%
    \put(0.35708922,0.05972222){\color[rgb]{0,0,0}\makebox(0,0)[t]{\lineheight{1.25}\smash{\begin{tabular}[t]{c}06:00\end{tabular}}}}%
    \put(0.46132828,0.05972222){\color[rgb]{0,0,0}\transparent{0}\makebox(0,0)[t]{\lineheight{1.25}\smash{\begin{tabular}[t]{c}09:00\end{tabular}}}}%
    \put(0.56556733,0.05972222){\color[rgb]{0,0,0}\makebox(0,0)[t]{\lineheight{1.25}\smash{\begin{tabular}[t]{c}12:00\end{tabular}}}}%
    \put(0.66980639,0.05972222){\color[rgb]{0,0,0}\transparent{0}\makebox(0,0)[t]{\lineheight{1.25}\smash{\begin{tabular}[t]{c}15:00\end{tabular}}}}%
    \put(0.77404544,0.05972222){\color[rgb]{0,0,0}\makebox(0,0)[t]{\lineheight{1.25}\smash{\begin{tabular}[t]{c}18:00\end{tabular}}}}%
    \put(0.8782845,0.05972222){\color[rgb]{0,0,0}\transparent{0}\makebox(0,0)[t]{\lineheight{1.25}\smash{\begin{tabular}[t]{c}21:00\end{tabular}}}}%
    \put(0,0){\includegraphics[width=\unitlength,page=2]{live555_auth_with_diff.pdf}}%
    \put(0.56527778,0.01805556){\color[rgb]{0,0,0}\makebox(0,0)[t]{\lineheight{1.25}\smash{\begin{tabular}[t]{c}\textbf{Time [hh:mm]}\end{tabular}}}}%
    \put(0,0){\includegraphics[width=\unitlength,page=3]{live555_auth_with_diff.pdf}}%
    \put(0.12916667,0.09305556){\color[rgb]{0,0,0}\makebox(0,0)[rt]{\lineheight{1.25}\smash{\begin{tabular}[t]{r}0\end{tabular}}}}%
    \put(0.12916667,0.24320569){\color[rgb]{0,0,0}\makebox(0,0)[rt]{\lineheight{1.25}\smash{\begin{tabular}[t]{r}500\end{tabular}}}}%
    \put(0.12916667,0.39335586){\color[rgb]{0,0,0}\makebox(0,0)[rt]{\lineheight{1.25}\smash{\begin{tabular}[t]{r}1,000\end{tabular}}}}%
    \put(0.12916667,0.54350601){\color[rgb]{0,0,0}\makebox(0,0)[rt]{\lineheight{1.25}\smash{\begin{tabular}[t]{r}1,500\end{tabular}}}}%
    \put(0,0){\includegraphics[width=\unitlength,page=4]{live555_auth_with_diff.pdf}}%
    \put(0.04298774,0.37916667){\color[rgb]{0,0,0}\rotatebox{90}{\makebox(0,0)[t]{\lineheight{1.25}\smash{\begin{tabular}[t]{c}\textbf{\#Covered Branches}\end{tabular}}}}}%
    \put(0,0){\includegraphics[width=\unitlength,page=5]{live555_auth_with_diff.pdf}}%
    \put(0.71413382,0.67210343){\color[rgb]{0,0,0}\makebox(0,0)[lt]{\lineheight{1.25}\smash{\begin{tabular}[t]{l}enabled\end{tabular}}}}%
    \put(0,0){\includegraphics[width=\unitlength,page=6]{live555_auth_with_diff.pdf}}%
    \put(0.85624097,0.67210343){\color[rgb]{0,0,0}\makebox(0,0)[lt]{\lineheight{1.25}\smash{\begin{tabular}[t]{l}disabled\end{tabular}}}}%
    \put(0.6715171,0.7100474){\color[rgb]{0,0,0}\makebox(0,0)[lt]{\lineheight{1.25}\smash{\begin{tabular}[t]{l}\textbf{Authentication}\end{tabular}}}}%
    \put(0.15275065,0.71051737){\color[rgb]{0,0,0}\makebox(0,0)[lt]{\lineheight{1.25}\smash{\begin{tabular}[t]{l}\textbf{Fuzzer}\end{tabular}}}}%
    \put(0.53379911,0.74182069){\color[rgb]{0,0,0}\makebox(0,0)[t]{\lineheight{1.25}\smash{\begin{tabular}[t]{c}\textbf{\livefivefivefiveServer}\end{tabular}}}}%
    \put(0,0){\includegraphics[width=\unitlength,page=7]{live555_auth_with_diff.pdf}}%
    \put(0.18691752,0.67291755){\color[rgb]{0,0,0}\makebox(0,0)[lt]{\lineheight{1.25}\smash{\begin{tabular}[t]{l}\textbf{\toolname}\end{tabular}}}}%
    \put(0,0){\includegraphics[width=\unitlength,page=8]{live555_auth_with_diff.pdf}}%
    \put(0.3736947,0.67419181){\color[rgb]{0,0,0}\makebox(0,0)[lt]{\lineheight{1.25}\smash{\begin{tabular}[t]{l}\textbf{\sgfuzz}\end{tabular}}}}%
    \put(0,0){\includegraphics[width=\unitlength,page=9]{live555_auth_with_diff.pdf}}%
  \end{picture}%
\endgroup%

%% file: content/06_discussion.tex
\section{Discussion}\label{sec:discussion}
In the following, we discuss potential limitations of our approach and explore directions for future research in the area of fault injection in the context of network fuzzing.

\paragraph{Access Requirement.}
Ideally, we have access to both peers so that we can observe the behavior of the client and the server and instrument one of them to turn it into a weird peer. 
As such, our current setup cannot be used to test instant messenger clients such as WhatsApp or Apple's iMessage, given that we cannot instrument the server to turn it into a weird peer. 
A related challenge is that our prototype implementation requires access to the weird peer's source code so that we can instrument it, which may not be possible if the software is proprietary or closed-source.

\paragraph{Timing-Related Vulnerabilities.}
Our instrumentation and the fuzzing setup (\eg synchronization via the FuzzMediator) introduce some performance overhead. 
While this does not affect the testing process per se (i.e., we have successfully detected many types of memory safety violations), this overhead can affect the communication timing between the client and the server, potentially masking timing-related issues that \fuzztructionnet might miss.

\paragraph{Differential and Cross-Implementation Testing.}
Besides detecting classic memory safety violations, our approach could also be well-suited for testing differences in behavior between different implementations of the same protocol. 
This aspect is particularly crucial for protocols requiring strict adherence to their specifications to maintain strong security guarantees. 
Furthermore, this approach could be particularly effective in uncovering implicit constraints encoded within the program code (\eg because the specification is ambiguous).
Protocols such as TLS, with implementations such as OpenSSL, GnuTLS, and LibreSSL, are particularly sensitive to deviations from their defined standards that could potentially introduce side-channel and other vulnerabilities.
To implement differential testing, we could observe and compare the behavior of different implementations when interacting with a weird peer application that has been mutated using our approach.
By applying cross-implementation testing, where various clients are paired with different servers, we could expose and address hidden assumptions, ensuring the protocol functions are robust and secure across diverse environments.
This approach would allow us to systematically investigate and identify inconsistencies in protocol execution, revealing potential security weaknesses.

\paragraph{\generatorsc Configurations.}
Another interesting topic for future work is the comprehensive enumeration of different client and server configurations. At the moment, a particular fuzzing setup uses a fixed, user-defined configuration of the peers. While some protocol features can be unlocked by changing one of the peers, others require a manual effort (\eg configuring different types of public keys or en-/disabling options could be beneficial when testing a TLS implementation to cover more code). 
Future work could also adapt \fuzztructionnet to support multi-threaded \generator{}s. While supported in principle, scheduling introduced non-determinism that needs to be addressed. That said, most clients---that usually serve as \generator---are single-threaded.

%% file: content/07_related_work.tex
\section{Related Work}\label{sec:related_work}

Our work combines techniques from network application fuzzing and fault injection. As we test network applications that all speak specific protocols, our work is also related to protocol testing.

\paragraph{Network Application Fuzzing.}
Closest to our work are network fuzzers, including \sgfuzz~\cite{ba2022sgfuzz}, \stateafl~\cite{natella2022stateafl}, \aflnet~\cite{pham2020aflnet}, \bleem~\cite{luo2023bleem}, \peach~\cite{peach}, \boofuzz~\cite{boofuzz}, \snipuzz~\cite{feng2021snipuzz}, \snapfuzz~\cite{andronidis2022snapfuzz}, and \nyxnet~\cite{schumilo2022nyxnet}. Beyond these tools, there is a Windows version of \aflnet called \textsc{NetAFL}~\cite{netafl}. %
Fundamentally, all of these methods primarily use one of two strategies.
First, \emph{replay-based fuzzers} work by intercepting and replaying live network traffic, performing mutations to produce server-compatible messages that respect the underlying message format. 
Second, \emph{MitM fuzzers} actively intercept and modify live (plaintext) messages without the need for pre-recorded communication.
Our method differs from these paradigms by using fault injection, eliminating the dependence on pre-recorded traffic or in-transit message changes. 
This unique approach allows us to use either the client or the server for fuzzing by taking advantage of the peer's ability to synthesize messages on its own. 
This not only bypasses the complexity associated with understanding proprietary message structures or cryptographic roadblocks, but also improves the breadth and depth of our fuzzing capabilities by testing the target in a more dynamic and unpredictable way.

\paragraph{Fault Injection.}
Several fuzzers that use fault injection to test regular applications exist~\cite{bars2023fuzztruction, liu2021ifizz, jiang2020fifuzz,sharma2024fuzzerr}. For example, \fuzztruction~\cite{bars2023fuzztruction} targets specific application pairs, such as in encryption or compression, to test the handling of data formats by generating and consuming data across applications. 
More recently, \textsc{FuzzERR}~\cite{sharma2024fuzzerr} strategically inserts faults into API calls to evaluate the robustness of error handling.
These techniques have conceptual similarities with our approach.
However, they are narrowly tailored to the interaction of data formats and cannot account for the complexity of network applications.
In contrast, our tool is specifically designed to control and exploit the nuanced dynamics of network protocols, enabling effective testing of network applications beyond the scope of existing domain-specific fuzzers.

\paragraph{Protocol Testing.}
Given that all of our targets use specific protocols, our work relates to the area of \emph{protocol testing}~\cite{brostean2020protocolfuzzing,somorovsky2016tlsattacker,fiterau2020analysis}, which is essentially about verifying that protocols conform to their specifications, maintain interoperability, and are robust against various types of attacks~\cite{hierons2009using}.
For example, \textsc{TLS-Attacker}~\cite{somorovsky2016tlsattacker} is a flexible framework for evaluating the security features of TLS implementations based on a manual implementation of the TLS specification.
Crucially, our goal is \emph{not} to test the protocol itself, \ie the specification, or to find implementation flaws in the sense of protocol deviations. 
Instead, we focus on injecting faults in a protocol-agnostic way to test the robustness of network applications.
\fuzztructionnet cannot detect specification errors or implementation deviations unless they manifest themselves in memory corruption.

\paragraph{State Machine Inference.}
Orthogonal research on automata learning~\cite{pham2020aflnet,daniele2024survey,daniel2018statemachine,chlosta2021states,Steffen2011,deruiter2015psf} has focused on inferring the state machine of network protocols based on observed behaviors, a generally hard task.
These methods treat the application as a black box oracle and infer the structure of the state machine by analyzing the application's responses to various inputs.
In contrast, our approach is agnostic of the state machine, and we rely on the instrumentation of one of the communication peers.
\fuzztructionnet could be used together with existing state machine learning algorithms to infer the current state or state transitions more efficiently.

%% file: content/08_conclusion.tex
\section{Conclusion}\label{sec:conclusion}

Despite being a critical security boundary, network applications have enjoyed little attention in terms of fuzz testing. In this work, we have analyzed challenges preventing fuzzers from effectively and efficiently doing so, then proposed a fault injection-based approach to overcome them. Injecting faults allows us to strike a balance between sending unexpected input that stresses the target and adhering to the input structure to exercise deeper functionality. Our evaluation shows that fault injection is a promising approach, outperforming all state-of-the-art fuzzers regarding coverage and found bugs. Notably, our approach is the only network fuzzer to test both clients and servers.

%% file: content/A_appendix.tex
\begin{table*}[t]
    \centering
    \caption{Target configurations used during the evaluation.}%
    \label{tab:appendix:target_configuration}
    \begin{adjustbox}{max width=\textwidth}
    \begin{tabular}{lll}
        \toprule
        \textbf{Target} & \textbf{Configuration} \\
        \midrule
\dropbearServer & \verb|dropbear -p <address> -a -F -r ed25519.key -E -B| \\
\dropbearClient & \verb|dbclient user@<address> -yy -i ed25519.key pwd| \\
\dcmtkServer & \verb|dcmrecv  --config-file storescp.cfg Default -d <port>| \\
\dcmtkClient & \verb|dcmsend -aet YOU_AET -aec DCM4CHEE -d <host> <port> image-00000.dcm| \\
\gnutlsServer & \verb|gnutls-serv  -a -d 1000 --earlydata --x509certfile=cert.pem --x509keyfile=key.pem -b -p <port>| \\
\gnutlsClient & \verb|gnutls-cli <address> --rehandshake --starttls -d <port> -b| \\
\libresslServer & \verb|openssl s_server -key key.pem -cert cert.pem -accept <port> -www -naccept 1| \\
\libresslClient & \verb|openssl s_client -connect <address> -status| \\
\livefivefivefiveServer & \verb|testOnDemandRTSPServer <port>| \\
\livefivefivefiveClient & \verb|testRTSPClient rtsp://<address>/mp3AudioTest| \\
\mosquittoServer & \verb|mosquitto -p <port>| \\
\multirow{2}{*}{\mosquittoClient} & \verb|mosquitto_pub -h <address> -p <port> -t TOPIC -m m --will-topic wt --will-retain \| \\
                                  & \verb|--will-payload wp --insecure --will-qos 1 --repeat 10 -r| \\
\nginxServer & \verb|nginx -c nginx.conf| \\
\opensslServer & \verb|openssl s_server -key key.pem -cert cert.pem -accept 44330 -www -naccept 1| \\
\opensslClient & \verb|openssl s_client -connect localhost:44330 -reconnect| \\
\sambaServer & \verb|smbd  -s smb.conf -F -i| \\
\sambaClient & \verb|smbclient  -p <port> -L //<address>| \\
\apacheServer & \verb|httpd  -X -f httpd.conf| \\
\pjsipServer & \verb|siprtp -p 5080 -r 4010 --auto-quit --call-report -i <address> <remote-address> | \\
\pjsipClient & \verb|siprtp  -p <port> -r 4000 -i <address>| \\
\curlClient & \verb|curl -v --insecure --parallel <address>| \\
\ngtcptwoServer & \verb|wsslserver <address> <port> server.key server.cert| \\
\ngtcptwoClient & \verb|wsslclient  --exit-on-all-streams-close <address> <port>| \\
\opensshServer & \verb|sshd -f sshd_config -4 -D -d -r -p <port>| \\
\multirow{2}{*}{\opensshClient} & \verb|ssh user@<address> -p<port> -o StrictHostKeyChecking=no -i rsa.key -g -A -R 127.0.0.1:18888 \| \\
                                & \verb|-R 127.0.0.1:18889 -vvvvv -X pwd| \\
        \bottomrule
    \end{tabular}
    \end{adjustbox}
\end{table*}

\section{Target Configuration}\label{sec:appendix:target_configuration}

To enable reproducibility, we list the exact commandline parameters for our targets in Table~\ref{tab:appendix:target_configuration}.

\begin{figure}[tb]
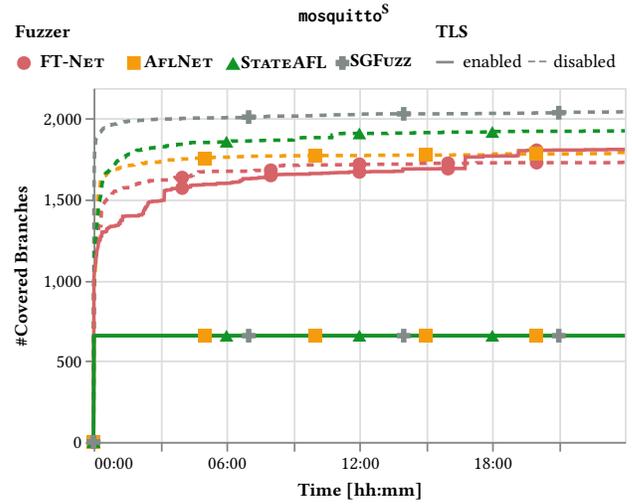

    \centering
    \graphicspath{{eval-plots/}}
    \def\svgwidth{\linewidth}
    \begin{footnotesize}
        
    \end{footnotesize}
    \caption{Coverage results for all fuzzers when enabling TLS support, as is often the case in practice. We compute coverage only for the target but not the TLS implementation.}%
    \label{fig:eval:mosquitto_tls_diff_all}
\end{figure}

\begin{figure}[t]
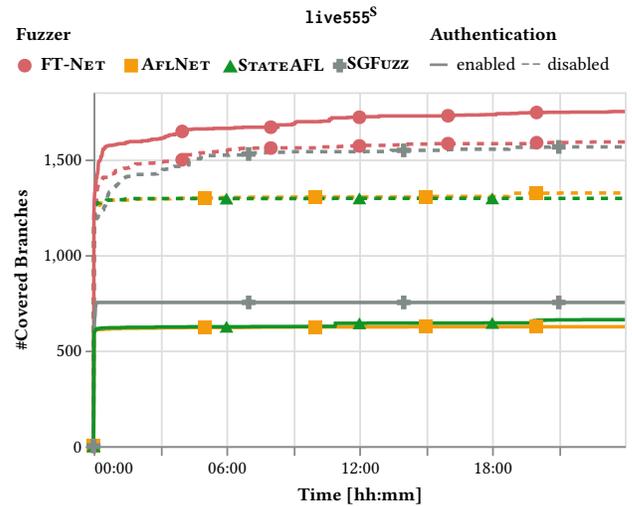

    \centering
    \graphicspath{{eval-plots/}}
    \def\svgwidth{\linewidth}
    \begin{footnotesize}
        
    \end{footnotesize}
    \caption{Coverage results when enabling HTTP Digest Access Authentication~\cite{rfc7616} for \livefivefivefiveServer. We provided all fuzzers with a recording of one successful authentication (HTTP plain-text).}%
    \label{fig:eval:live555_auth_diff_all}
\end{figure}

\section{Mosquitto and Live555}
When enabling TLS support for \mosquittoServer, only \fuzztructionnet manages to effectively explore it. In Figure~\ref{fig:eval:mosquito_tls}, we compared \fuzztructionnet only to \sgfuzz for readability reasons, and we provide the full data in Figure~\ref{fig:eval:mosquitto_tls_diff_all}. We observe similar results for \livefivefivefiveServer when using HTTP basic authentication, with the comparison of \fuzztructionnet and \sgfuzz in Figure~\ref{fig:eval:live555_auth} and the full data in Figure~\ref{fig:eval:live555_auth_diff_all}.

%% file: eval-plots/mosquitto_tls_with_diff_all.pdf_tex
\begingroup%
  \makeatletter%
  \providecommand\color[2][]{%
    \errmessage{(Inkscape) Color is used for the text in Inkscape, but the package 'color.sty' is not loaded}%
    \renewcommand\color[2][]{}%
  }%
  \providecommand\transparent[1]{%
    \errmessage{(Inkscape) Transparency is used (non-zero) for the text in Inkscape, but the package 'transparent.sty' is not loaded}%
    \renewcommand\transparent[1]{}%
  }%
  \providecommand\rotatebox[2]{#2}%
  \newcommand*\fsize{\dimexpr\f@size pt\relax}%
  \newcommand*\lineheight[1]{\fontsize{\fsize}{#1\fsize}\selectfont}%
  \ifx\svgwidth\undefined%
    \setlength{\unitlength}{270bp}%
    \ifx\svgscale\undefined%
      \relax%
    \else%
      \setlength{\unitlength}{\unitlength * \real{\svgscale}}%
    \fi%
  \else%
    \setlength{\unitlength}{\svgwidth}%
  \fi%
  \global\let\svgwidth\undefined%
  \global\let\svgscale\undefined%
  \makeatother%
  \begin{picture}(1,0.85)%
    \lineheight{1}%
    \setlength\tabcolsep{0pt}%
    \put(0,0){\includegraphics[width=\unitlength,page=1]{mosquitto_tls_with_diff_all.pdf}}%
    \put(0.14861111,0.05972222){\color[rgb]{0,0,0}\makebox(0,0)[lt]{\lineheight{1.25}\smash{\begin{tabular}[t]{l}00:00\end{tabular}}}}%
    \put(0.25285017,0.05972222){\color[rgb]{0,0,0}\transparent{0}\makebox(0,0)[t]{\lineheight{1.25}\smash{\begin{tabular}[t]{c}03:00\end{tabular}}}}%
    \put(0.35708923,0.05972222){\color[rgb]{0,0,0}\makebox(0,0)[t]{\lineheight{1.25}\smash{\begin{tabular}[t]{c}06:00\end{tabular}}}}%
    \put(0.46132829,0.05972222){\color[rgb]{0,0,0}\transparent{0}\makebox(0,0)[t]{\lineheight{1.25}\smash{\begin{tabular}[t]{c}09:00\end{tabular}}}}%
    \put(0.56556731,0.05972222){\color[rgb]{0,0,0}\makebox(0,0)[t]{\lineheight{1.25}\smash{\begin{tabular}[t]{c}12:00\end{tabular}}}}%
    \put(0.66980637,0.05972222){\color[rgb]{0,0,0}\transparent{0}\makebox(0,0)[t]{\lineheight{1.25}\smash{\begin{tabular}[t]{c}15:00\end{tabular}}}}%
    \put(0.77404548,0.05972222){\color[rgb]{0,0,0}\makebox(0,0)[t]{\lineheight{1.25}\smash{\begin{tabular}[t]{c}18:00\end{tabular}}}}%
    \put(0.87828454,0.05972222){\color[rgb]{0,0,0}\transparent{0}\makebox(0,0)[t]{\lineheight{1.25}\smash{\begin{tabular}[t]{c}21:00\end{tabular}}}}%
    \put(0,0){\includegraphics[width=\unitlength,page=2]{mosquitto_tls_with_diff_all.pdf}}%
    \put(0.56527778,0.01805556){\color[rgb]{0,0,0}\makebox(0,0)[t]{\lineheight{1.25}\smash{\begin{tabular}[t]{c}\textbf{Time [hh:mm]}\end{tabular}}}}%
    \put(0,0){\includegraphics[width=\unitlength,page=3]{mosquitto_tls_with_diff_all.pdf}}%
    \put(0.12916667,0.09305556){\color[rgb]{0,0,0}\makebox(0,0)[rt]{\lineheight{1.25}\smash{\begin{tabular}[t]{r}0\end{tabular}}}}%
    \put(0.12916667,0.21995265){\color[rgb]{0,0,0}\makebox(0,0)[rt]{\lineheight{1.25}\smash{\begin{tabular}[t]{r}500\end{tabular}}}}%
    \put(0.12916667,0.34684978){\color[rgb]{0,0,0}\makebox(0,0)[rt]{\lineheight{1.25}\smash{\begin{tabular}[t]{r}1,000\end{tabular}}}}%
    \put(0.12916667,0.47374687){\color[rgb]{0,0,0}\makebox(0,0)[rt]{\lineheight{1.25}\smash{\begin{tabular}[t]{r}1,500\end{tabular}}}}%
    \put(0.12916667,0.60064401){\color[rgb]{0,0,0}\makebox(0,0)[rt]{\lineheight{1.25}\smash{\begin{tabular}[t]{r}2,000\end{tabular}}}}%
    \put(0,0){\includegraphics[width=\unitlength,page=4]{mosquitto_tls_with_diff_all.pdf}}%
    \put(0.04298774,0.37916667){\color[rgb]{0,0,0}\rotatebox{90}{\makebox(0,0)[t]{\lineheight{1.25}\smash{\begin{tabular}[t]{c}\textbf{\#Covered Branches}\end{tabular}}}}}%
    \put(0,0){\includegraphics[width=\unitlength,page=5]{mosquitto_tls_with_diff_all.pdf}}%
    \put(0.72692747,0.69055537){\color[rgb]{0,0,0}\makebox(0,0)[lt]{\lineheight{1.25}\smash{\begin{tabular}[t]{l}enabled\end{tabular}}}}%
    \put(0,0){\includegraphics[width=\unitlength,page=6]{mosquitto_tls_with_diff_all.pdf}}%
    \put(0.86903462,0.69102681){\color[rgb]{0,0,0}\makebox(0,0)[lt]{\lineheight{1.25}\smash{\begin{tabular}[t]{l}disabled\end{tabular}}}}%
    \put(0.68431074,0.73738873){\color[rgb]{0,0,0}\makebox(0,0)[lt]{\lineheight{1.25}\smash{\begin{tabular}[t]{l}\textbf{TLS}\end{tabular}}}}%
    \put(0.02441635,0.73759014){\color[rgb]{0,0,0}\makebox(0,0)[lt]{\lineheight{1.25}\smash{\begin{tabular}[t]{l}\textbf{Fuzzer}\end{tabular}}}}%
    \put(0,0){\includegraphics[width=\unitlength,page=7]{mosquitto_tls_with_diff_all.pdf}}%
    \put(0.5410372,0.76360646){\color[rgb]{0,0,0}\makebox(0,0)[t]{\lineheight{1.25}\smash{\begin{tabular}[t]{c}\textbf{\mosquittoServer}\end{tabular}}}}%
    \put(0.11413877,0.690578){\color[rgb]{0,0,0}\makebox(0,0)[t]{\lineheight{1.25}\smash{\begin{tabular}[t]{c}\textbf{\toolname}\end{tabular}}}}%
    \put(0,0){\includegraphics[width=\unitlength,page=8]{mosquitto_tls_with_diff_all.pdf}}%
    \put(0.28032118,0.690578){\color[rgb]{0,0,0}\makebox(0,0)[t]{\lineheight{1.25}\smash{\begin{tabular}[t]{c}\textbf{\aflnet}\end{tabular}}}}%
    \put(0,0){\includegraphics[width=\unitlength,page=9]{mosquitto_tls_with_diff_all.pdf}}%
    \put(0.44589918,0.69002851){\color[rgb]{0,0,0}\makebox(0,0)[t]{\lineheight{1.25}\smash{\begin{tabular}[t]{c}\textbf{\stateafl}\end{tabular}}}}%
    \put(0,0){\includegraphics[width=\unitlength,page=10]{mosquitto_tls_with_diff_all.pdf}}%
    \put(0.59805076,0.69185227){\color[rgb]{0,0,0}\makebox(0,0)[t]{\lineheight{1.25}\smash{\begin{tabular}[t]{c}\textbf{\sgfuzz}\end{tabular}}}}%
    \put(0,0){\includegraphics[width=\unitlength,page=11]{mosquitto_tls_with_diff_all.pdf}}%
  \end{picture}%
\endgroup%

%% file: eval-plots/live555_auth_with_diff_all.pdf_tex
\begingroup%
  \makeatletter%
  \providecommand\color[2][]{%
    \errmessage{(Inkscape) Color is used for the text in Inkscape, but the package 'color.sty' is not loaded}%
    \renewcommand\color[2][]{}%
  }%
  \providecommand\transparent[1]{%
    \errmessage{(Inkscape) Transparency is used (non-zero) for the text in Inkscape, but the package 'transparent.sty' is not loaded}%
    \renewcommand\transparent[1]{}%
  }%
  \providecommand\rotatebox[2]{#2}%
  \newcommand*\fsize{\dimexpr\f@size pt\relax}%
  \newcommand*\lineheight[1]{\fontsize{\fsize}{#1\fsize}\selectfont}%
  \ifx\svgwidth\undefined%
    \setlength{\unitlength}{270bp}%
    \ifx\svgscale\undefined%
      \relax%
    \else%
      \setlength{\unitlength}{\unitlength * \real{\svgscale}}%
    \fi%
  \else%
    \setlength{\unitlength}{\svgwidth}%
  \fi%
  \global\let\svgwidth\undefined%
  \global\let\svgscale\undefined%
  \makeatother%
  \begin{picture}(1,0.85)%
    \lineheight{1}%
    \setlength\tabcolsep{0pt}%
    \put(0,0){\includegraphics[width=\unitlength,page=1]{live555_auth_with_diff_all.pdf}}%
    \put(0.14861111,0.05972222){\color[rgb]{0,0,0}\makebox(0,0)[lt]{\lineheight{1.25}\smash{\begin{tabular}[t]{l}00:00\end{tabular}}}}%
    \put(0.25285017,0.05972222){\color[rgb]{0,0,0}\transparent{0}\makebox(0,0)[t]{\lineheight{1.25}\smash{\begin{tabular}[t]{c}03:00\end{tabular}}}}%
    \put(0.35708922,0.05972222){\color[rgb]{0,0,0}\makebox(0,0)[t]{\lineheight{1.25}\smash{\begin{tabular}[t]{c}06:00\end{tabular}}}}%
    \put(0.46132828,0.05972222){\color[rgb]{0,0,0}\transparent{0}\makebox(0,0)[t]{\lineheight{1.25}\smash{\begin{tabular}[t]{c}09:00\end{tabular}}}}%
    \put(0.56556733,0.05972222){\color[rgb]{0,0,0}\makebox(0,0)[t]{\lineheight{1.25}\smash{\begin{tabular}[t]{c}12:00\end{tabular}}}}%
    \put(0.66980639,0.05972222){\color[rgb]{0,0,0}\transparent{0}\makebox(0,0)[t]{\lineheight{1.25}\smash{\begin{tabular}[t]{c}15:00\end{tabular}}}}%
    \put(0.77404544,0.05972222){\color[rgb]{0,0,0}\makebox(0,0)[t]{\lineheight{1.25}\smash{\begin{tabular}[t]{c}18:00\end{tabular}}}}%
    \put(0.8782845,0.05972222){\color[rgb]{0,0,0}\transparent{0}\makebox(0,0)[t]{\lineheight{1.25}\smash{\begin{tabular}[t]{c}21:00\end{tabular}}}}%
    \put(0,0){\includegraphics[width=\unitlength,page=2]{live555_auth_with_diff_all.pdf}}%
    \put(0.56527778,0.01805556){\color[rgb]{0,0,0}\makebox(0,0)[t]{\lineheight{1.25}\smash{\begin{tabular}[t]{c}\textbf{Time [hh:mm]}\end{tabular}}}}%
    \put(0,0){\includegraphics[width=\unitlength,page=3]{live555_auth_with_diff_all.pdf}}%
    \put(0.12916667,0.09305556){\color[rgb]{0,0,0}\makebox(0,0)[rt]{\lineheight{1.25}\smash{\begin{tabular}[t]{r}0\end{tabular}}}}%
    \put(0.12916667,0.24320569){\color[rgb]{0,0,0}\makebox(0,0)[rt]{\lineheight{1.25}\smash{\begin{tabular}[t]{r}500\end{tabular}}}}%
    \put(0.12916667,0.39335586){\color[rgb]{0,0,0}\makebox(0,0)[rt]{\lineheight{1.25}\smash{\begin{tabular}[t]{r}1,000\end{tabular}}}}%
    \put(0.12916667,0.54350601){\color[rgb]{0,0,0}\makebox(0,0)[rt]{\lineheight{1.25}\smash{\begin{tabular}[t]{r}1,500\end{tabular}}}}%
    \put(0,0){\includegraphics[width=\unitlength,page=4]{live555_auth_with_diff_all.pdf}}%
    \put(0.04298774,0.37916667){\color[rgb]{0,0,0}\rotatebox{90}{\makebox(0,0)[t]{\lineheight{1.25}\smash{\begin{tabular}[t]{c}\textbf{\#Covered Branches}\end{tabular}}}}}%
    \put(0,0){\includegraphics[width=\unitlength,page=5]{live555_auth_with_diff_all.pdf}}%
    \put(0.71711295,0.69312751){\color[rgb]{0,0,0}\makebox(0,0)[lt]{\lineheight{1.25}\smash{\begin{tabular}[t]{l}enabled\end{tabular}}}}%
    \put(0,0){\includegraphics[width=\unitlength,page=6]{live555_auth_with_diff_all.pdf}}%
    \put(0.8592201,0.69359896){\color[rgb]{0,0,0}\makebox(0,0)[lt]{\lineheight{1.25}\smash{\begin{tabular}[t]{l}disabled\end{tabular}}}}%
    \put(0.67449621,0.73996087){\color[rgb]{0,0,0}\makebox(0,0)[lt]{\lineheight{1.25}\smash{\begin{tabular}[t]{l}\textbf{Authentication}\end{tabular}}}}%
    \put(0.02571294,0.74016229){\color[rgb]{0,0,0}\makebox(0,0)[lt]{\lineheight{1.25}\smash{\begin{tabular}[t]{l}\textbf{Fuzzer}\end{tabular}}}}%
    \put(0,0){\includegraphics[width=\unitlength,page=7]{live555_auth_with_diff_all.pdf}}%
    \put(0.53677822,0.7661786){\color[rgb]{0,0,0}\makebox(0,0)[t]{\lineheight{1.25}\smash{\begin{tabular}[t]{c}\textbf{\livefivefivefiveServer}\end{tabular}}}}%
    \put(0.11543536,0.69315014){\color[rgb]{0,0,0}\makebox(0,0)[t]{\lineheight{1.25}\smash{\begin{tabular}[t]{c}\textbf{\toolname}\end{tabular}}}}%
    \put(0,0){\includegraphics[width=\unitlength,page=8]{live555_auth_with_diff_all.pdf}}%
    \put(0.27606221,0.69315014){\color[rgb]{0,0,0}\makebox(0,0)[t]{\lineheight{1.25}\smash{\begin{tabular}[t]{c}\textbf{\aflnet}\end{tabular}}}}%
    \put(0,0){\includegraphics[width=\unitlength,page=9]{live555_auth_with_diff_all.pdf}}%
    \put(0.44164022,0.69260065){\color[rgb]{0,0,0}\makebox(0,0)[t]{\lineheight{1.25}\smash{\begin{tabular}[t]{c}\textbf{\stateafl}\end{tabular}}}}%
    \put(0,0){\includegraphics[width=\unitlength,page=10]{live555_auth_with_diff_all.pdf}}%
    \put(0.59379181,0.69442441){\color[rgb]{0,0,0}\makebox(0,0)[t]{\lineheight{1.25}\smash{\begin{tabular}[t]{c}\textbf{\sgfuzz}\end{tabular}}}}%
    \put(0,0){\includegraphics[width=\unitlength,page=11]{live555_auth_with_diff_all.pdf}}%
  \end{picture}%
\endgroup%

%% file: main.bbl
\begin{thebibliography}{10}

\bibitem{andronidis2022snapfuzz}
Anastasios Andronidis and Cristian Cadar.
\newblock {Snapfuzz: High-throughput fuzzing of network applications}.
\newblock In {\em International Symposium on Software Testing and Analysis
  (ISSTA)}, 2022.

\bibitem{aschermann2019redqueen}
Cornelius Aschermann, Sergej Schumilo, Tim Blazytko, Robert Gawlik, and
  Thorsten Holz.
\newblock {REDQUEEN: Fuzzing with Input-to-State Correspondence}.
\newblock In {\em Symposium on Network and Distributed System Security (NDSS)},
  2019.

\bibitem{ba2022sgfuzz}
Jinsheng Ba, Marcel Böhme, Zahra Mirzamomen, and Abhik Roychoudhury.
\newblock {Stateful Greybox Fuzzing}.
\newblock In {\em USENIX Security Symposium}, 2022.

\bibitem{bars2023fuzztruction}
Nils Bars, Moritz Schloegel, Tobias Scharnowski, Nico Schiller, and Thorsten
  Holz.
\newblock {Fuzztruction: Using Fault Injection-based Fuzzing to Leverage
  Implicit Domain Knowledge}.
\newblock In {\em USENIX Security Symposium}, 2023.

\bibitem{Bratus2011weird}
Sergej Bratus, Michael~E. Locasto, Meredith~L. Patterson, Len Sassaman, and
  Anna Shubina.
\newblock {Exploit Programming: From Buffer Overflows to ``Weird Machines'''
  and Theory of Computation}.
\newblock {\em Usenix; Login}, 2011.

\bibitem{boehme2020entropic}
Marcel Böhme, Valentin J.~M. Manès, and Sang~Kil Cha.
\newblock {Boosting Fuzzer Efficiency: An Information Theoretic Perspective}.
\newblock In {\em ACM Joint European Software Engineering Conference and
  Symposium on the Foundations of Software Engineering (ESEC/FSE)}, 2020.

\bibitem{boehme2022_covreliability}
Marcel Böhme, László Szekeres, and Jonathan Metzman.
\newblock {On the Reliability of Coverage-Based Fuzzer Benchmarking}.
\newblock In {\em ACM/IEEE International Conference on Automated Software
  Engineering (ASE)}, 2022.

\bibitem{chlosta2021states}
Merlin Chlosta, David Rupprecht, and Thorsten Holz.
\newblock {On the Challenges of Automata Reconstruction in LTE Networks}.
\newblock In {\em ACM Conference on Security and Privacy in Wireless and Mobile
  Networks (WiSec)}, 2021.

\bibitem{daniel2018statemachine}
Lesly-Ann Daniel, Erik Poll, and Joeri de~Ruiter.
\newblock {Inferring OpenVPN State Machines Using Protocol State Fuzzing}.
\newblock In {\em IEEE European Symposium on Security and Privacy Workshops
  (EuroS\&PW)}, 2018.

\bibitem{daniele2024survey}
Cristian Daniele, Seyed~Behnam Andarzian, and Erik Poll.
\newblock {Fuzzers for Stateful Systems: Survey and Research Directions}.
\newblock {\em ACM Computing Surveys (CSUR)}, 56(9), 2024.

\bibitem{deruiter2015psf}
Joeri De~Ruiter and Erik Poll.
\newblock {Protocol State Fuzzing of TLS Implementations}.
\newblock In {\em USENIX Security Symposium}, 2015.

\bibitem{peach}
Michael Eddington.
\newblock {Peach Fuzzer: Discover Unknown Vulnerabilities}.
\newblock \url{https://peachtech.gitlab.io/peach-fuzzer-community/}, 2004.

\bibitem{feng2021snipuzz}
Xiaotao Feng, Ruoxi Sun, Xiaogang Zhu, Minhui Xue, Sheng Wen, Dongxi Liu, Surya
  Nepal, and Yang Xiang.
\newblock {Snipuzz: Black-box Fuzzing of IoT Firmware via Message Snippet
  Inference}.
\newblock In {\em ACM Conference on Computer and Communications Security
  (CCS)}, 2021.

\bibitem{aflplusplus}
Andrea Fioraldi, Dominik Maier, Heiko Ei{\ss}feldt, and Marc Heuse.
\newblock {AFL++ : Combining Incremental Steps of Fuzzing Research}.
\newblock In {\em USENIX Workshop on Offensive Technologies (WOOT)}, 2020.

\bibitem{fioraldi2022libafl}
Andrea Fioraldi, Dominik~Christian Maier, Dongjia Zhang, and Davide Balzarotti.
\newblock {LibAFL: A Framework to Build Modular and Reusable Fuzzers}.
\newblock In {\em ACM Conference on Computer and Communications Security
  (CCS)}, 2022.

\bibitem{brostean2020protocolfuzzing}
Paul Fiterau-Brostean, Bengt Jonsson, Robert Merget, Joeri de~Ruiter,
  Konstantinos Sagonas, and Juraj Somorovsky.
\newblock {Analysis of DTLS Implementations Using Protocol State Fuzzing}.
\newblock In {\em USENIX Security Symposium}, 2020.

\bibitem{fiterau2020analysis}
Paul Fiterau-Brostean, Bengt Jonsson, Robert Merget, Joeri De~Ruiter,
  Konstantinos Sagonas, and Juraj Somorovsky.
\newblock {Analysis of DTLS Implementations Using Protocol State Fuzzing}.
\newblock In {\em USENIX Security Symposium}, 2020.

\bibitem{honggfuzz}
{Google}.
\newblock {Honggfuzz}.

\bibitem{ossfuzz}
{Google}.
\newblock {OSS-Fuzz: Continuous Fuzzing for Open Source Software}.

\bibitem{google2016fts}
{Google}.
\newblock {Fuzzer-Test-Suite}, 2016.

\bibitem{hazimeh2020magma}
Ahmad Hazimeh, Adrian Herrera, and Mathias Payer.
\newblock {Magma: A Ground-Truth Fuzzing Benchmark}.
\newblock {\em ACM on Measurement and Analysis of Computing Systems (POMACS)},
  4(3):49:1--49:29, 2020.

\bibitem{hierons2009using}
Robert~M Hierons, Kirill Bogdanov, Jonathan~P Bowen, Rance Cleaveland, John
  Derrick, Jeremy Dick, Marian Gheorghe, Mark Harman, Kalpesh Kapoor, Paul
  Krause, et~al.
\newblock {Using Formal Specifications to Support Testing}.
\newblock {\em ACM Computing Surveys (CSUR)}, 41(2), 2009.

\bibitem{jiang2020fifuzz}
Zu-Ming Jiang, Jia-Ju Bai, Kangjie Lu, and Shi-Min Hu.
\newblock {Fuzzing Error Handling Code using Context-Sensitive Software Fault
  Injection}.
\newblock In {\em USENIX Security Symposium}, 2020.

\bibitem{boofuzz}
{Joshua Pereyda et al.}
\newblock {boofuzz: Network Protocol Fuzzing for Humans}.
\newblock \url{https://github.com/jtpereyda/boofuzz}.

\bibitem{klees2018evaluating}
George Klees, Andrew Ruef, Benji Cooper, Shiyi Wei, and Michael Hicks.
\newblock {Evaluating Fuzz Testing}.
\newblock In {\em ACM Conference on Computer and Communications Security
  (CCS)}, 2018.

\bibitem{lafintel}
{lafintel}.
\newblock {laf-intel - Circumventing Fuzzing Roadblocks with Compiler
  Transformations}.
\newblock \url{https://lafintel.wordpress.com}.

\bibitem{liu2021ifizz}
Peiyu Liu, Shouling Ji, Xuhong Zhang, Qinming Dai, Kangjie Lu, Lirong Fu,
  Wenzhi Chen, Peng Cheng, Wenhai Wang, and Raheem Beyah.
\newblock {IFIZZ: Deep-State and Efficient Fault-Scenario Generation to Test
  IoT Firmware}.
\newblock In {\em ACM/IEEE International Conference on Automated Software
  Engineering (ASE)}, 2021.

\bibitem{luo2023bleem}
Zhengxiong Luo, Junze Yu, Feilong Zuo, Jianzhong Liu, Yu~Jiang, Ting Chen,
  Abhik Roychoudhury, and Jiaguang Sun.
\newblock {Bleem: Packet Sequence Oriented Fuzzing for Protocol
  Implementations}.
\newblock In {\em USENIX Security Symposium}, 2023.

\bibitem{lyu2019mopt}
Chenyang Lyu, Shouling Ji, Chao Zhang, Yuwei Li, Wei-Han Lee, Yu~Song, and
  Raheem Beyah.
\newblock {MOPT: Optimized Mutation Scheduling for Fuzzers}.
\newblock In {\em USENIX Security Symposium}, 2019.

\bibitem{metzman2021fuzzbench}
Jonathan Metzman, László Szekeres, Laurent Simon, Read Sprabery, and Abhishek
  Arya.
\newblock {FuzzBench: An Open Fuzzer Benchmarking Platform and Service}.
\newblock In {\em ACM Joint European Software Engineering Conference and
  Symposium on the Foundations of Software Engineering (ESEC/FSE)}, 2021.

\bibitem{natella2022stateafl}
Roberto Natella.
\newblock {StateAFL: Greybox Fuzzing for Stateful Network Servers}.
\newblock {\em Empirical Software Engineering}, 27(7):191, 2022.

\bibitem{natella2021profuzzbench}
Roberto Natella and Van-Thuan Pham.
\newblock {ProFuzzBench: A Benchmark for Stateful Protocol Fuzzing}.
\newblock In {\em International Symposium on Software Testing and Analysis
  (ISSTA)}, 2021.

\bibitem{pham2020aflnet}
Van-Thuan Pham, Marcel B{\"o}hme, and Abhik Roychoudhury.
\newblock {AFLNet: A Greybox Fuzzer for Network Protocols}.
\newblock In {\em IEEE International Conference on Software Testing, Validation
  and Verification (ICST)}, 2020.

\bibitem{schloegel2024sok}
Moritz Schloegel, Nils Bars, Nico Schiller, Lukas Bernhard, Tobias Scharnowski,
  Addison Crump, Arash Ale-Ebrahim, Nicolai Bissantz, Marius Muench, and
  Thorsten Holz.
\newblock {SoK: Prudent Evaluation Practices for Fuzzing}.
\newblock In {\em IEEE Symposium on Security and Privacy (S\&P)}, 2024.

\bibitem{schumilo2022nyxnet}
Sergej Schumilo, Cornelius Aschermann, Andrea Jemmett, Ali Abbasi, and Thorsten
  Holz.
\newblock {Nyx-Net: Network Fuzzing with Incremental Snapshots}.
\newblock In {\em European Conference on Computer Systems (EuroSys)}, 2022.

\bibitem{sharma2024fuzzerr}
Shashank Sharma, Sai~Ritvik Tanksalkar, Sourag Cherupattamoolayil, and Aravind
  Machiry.
\newblock {Fuzzing API Error Handling Behaviors using Coverage Guided Fault
  Injection}.
\newblock In {\em ACM Symposium on Information, Computer and Communications
  Security (ASIACCS)}, 2024.

\bibitem{rfc7616}
Rifaat Shekh-Yusef.
\newblock {HTTP Digest Access Authentication}, 2015.

\bibitem{netafl}
M.~Shudrak.
\newblock {NetAFL: WinAFL Patch}.
\newblock \url{https://github.com/intelpt/winafl-intelpt}, 2018.

\bibitem{somorovsky2016tlsattacker}
Juraj Somorovsky.
\newblock {Systematic Fuzzing and Testing of TLS Libraries}.
\newblock In {\em ACM Conference on Computer and Communications Security
  (CCS)}, 2016.

\bibitem{Steffen2011}
Bernhard Steffen, Falk Howar, and Maik Merten.
\newblock {\em {Introduction to Active Automata Learning from a Practical
  Perspective}}.
\newblock 2011.

\bibitem{vargha2000critique}
Andr{\'a}s Vargha and Harold~D Delaney.
\newblock {A Critique and Improvement of the CL Common Language Effect Size
  Statistics of McGraw and Wong}.
\newblock {\em Journal of Educational and Behavioral Statistics},
  25(2):101--132, 2000.

\bibitem{yun2018qsym}
Insu Yun, Sangho Lee, Meng Xu, Yeongjin Jang, and Taesoo Kim.
\newblock {QSYM: A Practical Concolic Execution Engine Tailored for Hybrid
  Fuzzing}.
\newblock In {\em USENIX Security Symposium}, 2018.

\bibitem{afl}
Micha\l{} Zalewski.
\newblock {American Fuzzy Lop}.
\newblock \url{https://lcamtuf.coredump.cx/afl/}, 2013.

\bibitem{zhao2022statefuzz}
Bodong Zhao, Zheming Li, Shisong Qin, Zheyu Ma, Ming Yuan, Wenyu Zhu, Zhihong
  Tian, and Chao Zhang.
\newblock {StateFuzz: System Call-Based State-Aware Linux Driver Fuzzing}.
\newblock In {\em USENIX Security Symposium}, 2022.

\bibitem{zhu2022survey}
Xiaogang Zhu, Sheng Wen, Seyit Camtepe, and Yang Xiang.
\newblock {Fuzzing: A Survey for Roadmap}.
\newblock {\em ACM Computing Surveys (CSUR)}, 54(11s):1--36, 2022.

\end{thebibliography}
